\documentclass[11pt]{article}

\usepackage{graphicx}
\usepackage{graphicx,psfrag}
\usepackage{amsmath,amssymb,amsthm}

\def \G {\Gamma}

\def \E {{\bf  \, E  }}

\def \Tr {{\mathrm{\, Tr}}}

\def \bC {{\mathbb C}}

\def \bD {{\mathbb D}}

\def \bI {{\mathbb I}}

\def \bJ {{\mathbb J}}

\def \bN {{\mathbb N}}

\def \bR {{\mathbb R}}

\def \bU {{\mathbb U}}

\def \bX {{\mathbb X}}

\def \bW {{\mathbb W}}

\def \bY {{\mathbb Y}}

\def \bZ {{\mathbb Z}}

\def \CA {{\cal A}}

\def \CE {{\cal E}}
\def \CC {{\cal C}}

\def \CG {{\cal G}}
\def \CH {{\cal H}}
\def \CU {{\cal U}}
\def \CN {{\cal N}}
\def \CT {{\cal T}}

\def \CW {{\cal W}}
\def \CD {{\cal D}}
\def \CI {{\cal I}}
\def\CL {{\cal L}}
\def \CQ {{\cal Q}}

\def \CR {{\cal R}}

\def \CJ {{\cal J}}

\def \CF {{\cal F}}

\def \CS {{\cal S}}

\def \CV {{\cal V}}

\def \G{{\Gamma}}

\def \a {\alpha}

\def \b {\beta}
\def \g {\gamma}
\def \l {\lambda}
\def \r {\rho}

\def \s {\sigma}
\def \t {\tau}

\def \d {\delta}
\def \U {\Upsilon}
\def \u {\upsilon}
\def \th {\theta}

\def \Th {\Theta}

\def \vep {\varepsilon}

\def \ep {\epsilon}

\def \L {\Lambda}

\def \fb {{\mathfrak b}}

\def \fI {{\mathfrak I}}

\def \fL {{\mathfrak L}}

\def \fr {{\mathfrak r}}

\def \det {\hbox{det}}

\def \tA {\tilde A}
\def \tB {\tilde B}

\begin{document}
\title{  Limiting Eigenvalue Distribution of Random Matrices of  Ihara Zeta Function of  Long-Range Percolation  Graphs
}

\author{O. Khorunzhiy\\ Universit\'e de Versailles - Saint-Quentin \\45, Avenue des Etats-Unis, 78035 Versailles, FRANCE\\
{\it e-mail:} oleksiy.khorunzhiy@uvsq.fr}
\maketitle
\begin{abstract}
We consider  the ensemble of $N\times N$ 
real random symmetric matrices $H_N^{(R)}$ obtained 
from the determinant 
form of the Ihara zeta function associated to  random graphs $\Gamma_N^{(R)}$ 
of the long-range percolation radius model with the edge probability 
determined by a function $\phi(t)$.

We show that the normalized eigenvalue counting function   
 of $H_N^{( R)}$  weakly converges in average as $N,R\to\infty$, $R=o(N)$ to 
a unique  measure that depends  on the limiting 
average vertex degree of $\Gamma_N^{(R)}$ given by  
$\phi_1 = \int \phi(t) dt$. 
This measure converges in the limit of infinite $\phi_1$ to a shift of the Wigner semi-circle distribution. 
We discuss relations of these results  with  
the properties of the Ihara zeta function and weak versions of the  graph theory Riemann Hypothesis.

\end{abstract}
\vskip 0.3cm



\section{Introduction: Ihara zeta function and random matrices}

The Ihara zeta function introduced by Y. Ihara in the algebraic 
context \cite{I} attracts considerable interest also due to its 
interpretations in the frameworks of  graph theory. 
Given a finite connected non-oriented graph
\mbox{$\Gamma= (V,E)$} with the vertex set $V= \{\a_1,\a_2,\dots, \a_N\}$
and the edge set $E$, the Ihara zeta function (IZF) 
of $\Gamma$ 
 is defined for $u  \in \bC$ with $\vert u\vert $ sufficiently small,  
 by  
$$
Z_\G(u) = \prod _{[C]}\  (1- u^{\nu(C)})^{-1},
\eqno (1.1) 
$$
where the product runs over the equivalence classes of primitive closed backtrackless, tailless
 paths $C= (\a_{i_1},\a_{i_2}, \dots, \a_{i_{l-1}}, \a_{i_l})$, $\a_{i_l}=\a_{i_1}$ of positive length $l$  such that 
$\{\a_{i_h},\a_{i_{h+1}}\}\in E$
for all $h=1, \dots, l-1$ 
and \mbox{$\nu(C)=l-1$}  \cite{Te}. Here and below we denote by 
$e=\{\a_i,\a_j\}$ a non-oriented edge $e\in E$ if it exists.  
In this definition (1.1), a closed path
$C= (\a_{i_1},\a_{i_2}, \dots, \a_{i_{l-1}}, \a_{i_1})$ 
is primitive if there is no smaller path $\tilde C$
such  that $C=\tilde C^k$. The path $C$ is backtrackless if 
$\a_{i_{h-1}}\neq \a_{i_{h+1}}$
for all $h=2, \dots, l-1$. The closed path $C$ is tailless if 
$\a_{i_2}\neq \a_{i_{l-1}}$.  The equivalence class of closed paths $[C]$ includes $ C$ and all paths obtained with the help of the cyclic permutations of its elements, i. e.  of the form $C'=(\a_{i_2}, \dots, \a_{i_{l-1}},  \a_{i_1}, \a_{i_2})$, $C''=(\a_{i_3}, \dots, \a_{i_{l-1}}, \a_{i_1},\a_{i_2},\a_{i_3})$, etc.
Let us note that the closed primitive backtrackless tailless paths 
can be 
imagined as analogs of the closed geodesics on manifolds. 
In this sense the 
absence of backtracking can be considered as the absence of
singular (non-differentiable) points and the property to be tailless is analogous 
to differentiability of the geodesics at the origin.

The Ihara zeta function can also be expressed as 
$$
Z_\G (u) = \exp\left\{\sum_{k\ge 1} {\CN_k\over k} u^k \right\},
\eqno (1.2)
$$
where $\CN_k$ is the number of all classes of backtrackless tailless paths of the length $k$.
Ihara's theorem \cite{I}  says that the IZF (1.1)
is the reciprocal of a polynomial and that for sufficiently small 
$\vert u\vert$
$$
Z_\G(u)^{-1} = (1-u^2)^{r-1}\,  \det (I+u^2 (B-I) - uA),
\eqno (1.3)
$$
where $A = (a_{ij})_{i,j=1,\dots, N}$ is the adjacency matrix of $\G$, $N $ is the number of vertices, $N = \# V$, 
$B = diag( \sum_{j=1}^N a_{ij})_{i=1,\dots,N}$ and 
$r-1 = \Tr (B-2I)/2$.
 Relation (1.3) 
  proved  by Y. Ihara firstly for  the families of regular  graphs 
  that have the constant vertex degree, 
  $\deg(\a_i)= \deg(\a_j)$, $i,j=1, \dots, N$ 
  has been then  generalized  to the case of 
  possibly irregular finite graphs where the degrees of different vertices may be different (see e.g. papers \cite{Ba,ST} for the proofs on the basis of representation (1.2)).
 
At present, IZF is a well established part  of the combinatorial graph theory with applications 
in   number theory and  spectral theory
 (see e.g. \cite{HST,Te} and references therein); 
 it has also been studied in various other aspects, in particular 
 in relations with the heat kernels on graphs \cite{CJK}, 
 quantum walks on graphs \cite{RAE}, certain theoretical physics models 
 \cite{ZXH}. Relations between the Ihara zeta function and  spectral theory is a source of
 new interesting questions and problems.

  While the Ihara's determinant formula (1.3) gives a powerful tool in the studies of  the Ihara zeta function,
 the explicit form of $Z_\G(u)$ can be computed  
 for relatively narrow families of  finite graphs. 
The definition of the IZF  associated to infinite graphs requires 
 a number of additional  restrictions and assumptions
 (see, in particular, \cite{CMS,GIM-2,GZ,LPS}). 
 In the most studied cases, the graphs under consideration have a bounded vertex degree
 (in particular, regular or essentially regular graphs).

  A complementary approach  is related with the  probabilistic point of view,
 when  graphs are chosen at random from the family of all possible graphs on $N$ vertices.  
Such a description   naturally leads to the limiting transition  of infinitely increasing
 dimension of the graphs, $N\to\infty$.
 Certain aspects of zeta functions of
 large random regular graphs with finite vertex degree  have been studied in \cite{F}.

 Another class of random graphs is represented by the 
 Erd\H os-R\'enyi ones \cite {Bo}. 
 This family of graphs $\{\G_N^{(p)}\}$ can be described by the ensemble of real symmetric adjacency $N\times N$ matrices
 $A_N^{(p)}$
 whose entries above the diagonal are given by
  jointly independent Bernoulli random 
 variables with the average value $p$. In this model, the 
 vertex degree is a random variable with the average  value $pN$ that can be either finite or not in the limit $N\to\infty$. 

In paper \cite{K-15} we considered 
 the eigenvalue distribution of random matrix ensemble 
 obtained from the determinant formula (1.3) with the help of
 the adjacency matrices $A_N^{(\rho/N)}$ of the Erd\H os-R\'enyi random graphs. 
 It is  shown that after a proper renormalization
 of the spectral parameter $u$ by $\rho^{-1/2}$, this eigenvalue distribution 
 converges in the limit $N,\rho\to\infty$ when $ \rho=o(N^\alpha)$ for any  $\alpha >0$ to a shift of the widely known semi-circle Wigner distribution.

  In the present paper we consider similar questions
 for 
random matrices related with random  long-range 
percolation  radius graphs \cite{B,S,ZPL}, where the edge probability is a function of a kind of the distance between sites (vertices). 
These graphs  can be thought as a  generalization of the 
 Erd\H os-R\'enyi ensemble
that  give somewhat better 
description of certain systems  of interacting particles.
Our main observation  is that in the limit of infinite 
 graph dimension  and infinite interaction radius,
the moments of the limiting
eigenvalue distribution of the random matrix ensemble
generated by the determinant formula (1.3) 
for random long-range percolation graphs
verify a system of explicit relations  in the case when 
 the average vertex degree 
 $\phi_1$ remains finite.  Regarding the additional limiting
  transition $\phi_1\to\infty$, we 
  obtain convergence of the limiting distribution to 
  the same shifted semi-circle Wigner law 
  as described in \cite{K-15}.

\vskip 0.3cm

Let us consider the ensemble of $N\times N$  real symmetric matrices 
$A_N^{( R)}$,
$$
\left( A_N^{(R,\phi)}\right)_{xy}= a_{xy}^{(R,\phi)}, \quad \vert x\vert \le n, \  \vert y\vert  \le n, \ N=2n+1,\  R\ge 1 
\eqno (1.4)
$$ 
whose entries above the main diagonal 
are given   by a family of jointly independent Bernoulli random variables
$\CA^{(R,\phi)} = \{a^{(R,\phi)}_{xy}, \ x,y\in \bZ, \ x\le y  \}$,  such that
$$
a^{( R,\phi)}_{xy} = 
 \begin{cases}
  1-\delta_{x,y} , & \text{with probability ${ 1\over R} \phi\left({x-y\over R}\right) = p_R\, $} , \\
0, & \text {with probability $1-{p_R}\, $,}
\end{cases} 
\eqno (1.5)
$$
where  $\d_{x,y}$ is the Kronecker $\d$-symbol. 
Then the matrix obtained is symmetrized to get (1.4).
We assume that  $\phi(t), t\in\bR$ is a  real continuous  even  function 
such that $0<\phi(t)<1$ and 
$$
\phi_1= \int_{\bR} \phi(t) dt <+\infty \, .
\eqno (1.6)
$$ 
We assume for simplicity that $\phi(t)$ is 
  strictly decreasing for all   $t\ge 0$.
  In what follows, we omit the superscripts $R$ and $\phi$ in $A_N^{(R,\phi)}$,
  $\CA^{(R,\phi)}$ and $a_{xy}^{(R,\phi)}$ everywhere  when no confusion can arise.
  
The ensemble of random graphs $\{\G_N^{(R)}\}$ determined by  the adjacency matrices $A_N^{(R)}$ (1.3)
is    close to the well-known ensemble of random long-range 
 percolation radius graphs \cite{A,B,S, ZPL}. 
In certain cases one considers  graphs that have a number of non-random  edges. 
We do not add to the graphs the non-random edges and assume that
there are no loops in $\G$.  To generalize the existing models,    we introduce the   additional parameter  $R$  
into the definition of the edge probability $p_R$ that has not been used in the long-range percolation models cited above.

\vskip 0.1cm
We introduce the diagonal matrix 
$$
\left(B_N^{(R)}\right)_{xt} =  \d_{x,t}\ \sum_{y=-n}^n a^{(R)}_{xy}, \quad \vert x\vert, \vert t\vert \le n,
\eqno (1.7)
$$
and consider 
the logarithm of the Ihara zeta function of $\G_N^{(R)}$,  
$$
- {1\over N} \log Z_{\G_N^{(R)}}(u)= {1\over 2N}\Tr\left( B_N^{(R)}-2I\right) \log (1-u^2)  
$$
$$+ 
{1\over N}
\log \det \left(I(1-u^2)+u^2 B_N^{(R)} - u A_N^{(R)}\right).
\eqno (1.8)
$$
Regarding the first term of the right-hand side of (1.8)
$$
\Upsilon_N^{(R)} = {1\over 2N}\Tr( B_N^{(R)}-2I) \log (1-u^2),
$$
it is easy to compute its mathematical expectation 
 $\E \Upsilon_N^{(R)}$ with respect to the measure generated by the family $\CA^{(R)}$,
  $$
 \E \Upsilon_N^{(R)}= \left( {1\over 2NR} 
 \sum_{x,t=-n}^n \phi\left( {x-t\over R}\right)  -
  {1\over 2R}\phi(0) -1\right) \log (1-u^2).
 \eqno (1.9)
$$
In present paper we consider the asymptotic regime of large $N$ and $R$ 
$$
N\to\infty, \ R\to\infty, \ R=o(N)
\eqno (1.10)
$$
that we denote by $(N,R)^\circ\to\infty$. It follows from (1.6) and (1.9) that 
$$
\lim_{(N,R)^\circ\to\infty} \E \Upsilon_N^{(R)} = \left({\phi_1\over 2} -1 \right) \log (1-u^2).
\eqno (1.11)
$$
and therefore it is natural to consider a rescaling of the parameter $u$  given by
$$
u = {v\over  \sqrt{\phi_1}}.
\eqno (1.12)
$$
Let us note that $\phi_1$ represents the average vertex degree of $\G_N^{(R)}$ in the limit (1.10).

Taking into account (1.12), we rewrite  the last term   of (1.8)  as
$$
\Psi_N^{(R)} = {1\over N}
\log \det \left(\left( 1-{v^2\over \phi_1 } \right)I + H_N^{(R)}(v)\right), 
\eqno (1.13)
$$
 where
 $$
   H_N^{( R)}(v) =  {v^2\over \phi_1 } B_N^{(R)} - {v\over \sqrt {\phi_1}} A_N^{(R)} 
   = \tilde B_N^{(R)} - \tilde A_N^{(R)}.
 \eqno (1.14)
 $$ 
  The main goal of the present paper is to study the limiting eigenvalue distribution of these random matrices  with real $v$.
  Up to our knowledge, the random matrix ensemble $\{H_N^{(R)}(v)\}$ is a new one. It arises from the determinant formula for the Ihara zeta function (1.3) with a particular choice of the renormalization 
  of the spectral parameter (1.12) determined by the asymptotic expression of the factor 
  $(1-u)^{r-1}$. 
  As far as we know, the limiting spectral properties of  random matrices (1.14)   have not yet been    studied. 

  \section{Eigenvalue counting function of  random matrices}
  
  Let us denote the eigenvalues of the real symmetric matrix 
  $ H_N^{(R)}(v)$, $v\in \bR$  by $\l_1\le \dots\le  \l_N$. 
We consider 
the normalized eigenvalue counting function,
$$
\s_{N,R}(\l) = {1\over N} \sum_{j: \, \l_j\le \l} 1,
$$
and denote by   $\E \s_{N, R}= \bar \s_{N,R}$ its  average with respect to the measure generated by ${\cal A}^{(R)}$.
The measure   $\bar \s_{N,R}$ can be characterized by its  moments,
$$
M_{k}^{( N,R )} = \int_{\bR} \, \l^k d\bar \s_{N,R}(\l)
= {1\over N}\E  \Tr \left( H_N^{(R)}\right)^k.
\eqno (2.1)
$$
The main  result of the present paper is as follows.

\vskip 0.2cm
{\bf Theorem 2.1.} {\it 
Given $k\in \bN$, the  moment (2.1) converges in the limit (1.10)
to  $m_k^{(v,\phi_1)}$,
$$
\lim_{(N,R)^\circ\to\infty} M_k^{( N,R)} = m_k^{(v,\phi_1)}.
 \eqno (2.2) 
$$
This limiting moment  $m_k^{(v,\phi_1)}$ is given by relation
$$
m_k^{(v,\phi_1)} = \sum_{r=1}^k \Theta(k,r),
\eqno (2.3)
$$
where  the family $\{\Theta(k,r), \ 1\le r \le k\}_{ k\in \bN}$ is determined
by  the following recurrence,
$$
 \Th(k,r) =\  \sum_{g=1}^r\ \sum_{s=r-g}^{k-g} {r-1\choose g-1}\, \Th(s, r-g)
 $$
 $$
 \times\ 
 \sum_{w=0}^g\ \  \sum_{h=0}^{k-s-g-w}
 \ \ \sum_{t=0}^{k-s-g-w-h}\  \ {v^{2(g+h)}\over \phi_1^{g+h-1}} \ {g\choose w} \,  
 $$
 $$
 \times \ {w+h-1\choose  h}^* \  { w+h+t-1\choose t}^* \ \Th(k-s-g-w-h, t),
 \eqno (2.4)
 $$
  with the initial conditions 
$ \Th(k,0) = \d_{k,0}, \ k\ge 0.$
In (2.4),  the generalized binomial coefficient is such that 
$$
  {i-1\choose i}^* = \d_{i,0}, \quad i=0,1,2,\dots .
 \eqno (2.5)
$$}
\vskip 0.2cm 
It should be  pointed out  that  relations similar to (2.4) have been obtained 
in our previous work \cite{KSV}    (see also \cite{BG,KV}) 
for  similar but different ensemble of random matrices. 
In papers \cite{KSV,KV} we studied the limiting eigenvalue distribution of 
the  discrete analog of the Laplace operator of the Erd\H os-R\'enyi ensemble of random graphs.
This discrete analog has been normalized by $\rho^{-1/2}$, where 
$\rho/n$ represents the edge probability of the random $n$-dimensional graphs.  
Due to this difference in the normalization factor, random matrices $H_N^{(R)}(v)$ 
we consider here are no more positive definite.
In consequence,  relations (2.4)  differ
from those obtained in \cite{KSV} as well as the limiting eigenvalue distribution obtained after
the transitions $\phi_1\to\infty$ and  $\rho\to\infty$, respectively. 

\vskip 0.3cm

The mathematical expectation of the right-hand side of (2.1) 
 $$
\E \Tr(H^k) = {1\over N} \sum_{x_1, x_2, \dots, x_n}
\E\left( H_{x_1x_2} H_{x_2x_3}\cdots H_{x_{k-1}x_1}^{ }\right)
 $$ 
 can be considered as  
a sum over the set of 
weighted closed walks of $k$ steps $(x_1,\dots, x_{k-1},x_1)$.
This 
interpretation   proposed in  the pioneering works by E. Wigner (see e. g. \cite{W})  has been used, in one or another form, 
in a large  number of papers on random matrix theory.
We follow this strategy to study the product
$$
{\cal S}^{(N,R)}_{(\bar p, \bar q)}(s) ={1\over N} \sum_{x_1=-n }^n \cdots \sum_{x_s=-n }^n 
\ \E \left(  \tA^{p_1}_{x_1 x_2} \, \tB^{q_1}_{x_2x_2} \tA ^{p_2}_{x_2x_3}\cdots \tB^{q_{s-1}}_{x_sx_s}
\tA ^{p_s}_{x_s x_1} 
\tB ^{q_s}_{x_1x_1} \right) , 
\eqno (2.6)
$$ 
where the numbers $p_i, q_j$ are such that 
$$
p_1\ge 0, p_i>0, i=2,\dots, s, \ q_s\ge 0, q_j>0, j=1,\dots, s-1, 
\ 
P+Q=k,
$$
where $P=\sum_{i=1}^s p_i$ and $Q=\sum_{i=1}^s q_i$.
It is clear that 
we can rewrite (2.1) as follows,
$$
M_k^{(N,R)} = \sum_{s=1}^k \ \sum_{(\bar p, \bar q)^*}^k
(-1)^P \CS^{(N,R)}_{(\bar p, \bar q)}(s),
\eqno (2.7)
$$
where 
 $\bar p = (p_1, p_2,\dots, p_s)$, $\bar q = (q_1,q_2\dots,  q_s)$,
and the star in  the last sum means that it runs over all  
$(\bar p,\bar q)$ from (2.6).

\subsection{Color diagrams and weights}

Let us consider the  set of variables of the summation of (2.6)
$$
\bar x_{\bar p}^{(s)} = \left\{ \left(x_i^{(1)},\dots, x_i^{(p_i)}, x_i^{(p_i+1)}\right)\right\}_{i=1}^s
\ {\hbox{and}} \ \ 
\bar y_{\bar q}^{(s)} = \left\{ \left(y_j^{(1)}, \dots, y_j^{(q_j)}\right)\right\}_{j=1}^s,
$$
where $x^{(1)}_i = x_i$, $x_i^{(p_i+1)}= x_{i+1}= x_{i+1}^{(1)}$, $i=1, \dots , s$ and $x_s^{(p_s+1)}=x_{s+1} = x_1$ such that 
$$
\tilde A ^{p_i}_{x_i x_{i+1} } =
\left({v\over \sqrt {\phi_1}}\right)^{p_i}
\sum_{x_i^{(2)}, \dots, x_i^{(p_i)}}\ 
 a_{x_i^{(1)} x_i^{(2)}}\,
 a_{x_i^{(2)} x_i^{(3)}}\, \cdots \,
 a_{x_i^{(p_{i})} x_i^{(p_i+1)}}
$$
and 
$$
\tilde B^{q_j}_{x_{j+1}x_{j+1}}= 
\left({v^2\over {\phi_1}}\right)^{q_j}
\sum_{y_j^{(1)}, \dots, y_j^{(q_j)}}
a_{x_{j+1}y_j^{(1)}} \,
a_{x_{j+1}y_j^{(2)}}\,
\cdots\, 
a_{x_{j+1}y_j^{(q_j)}}, \quad j=1,\dots, s.  
$$\, 
Then can we rewrite the right-hand side of (2.6) as follows, 
$$
{1\over N} \sum_{x_1=-n }^n \cdots \sum_{x_s=-n }^n  \E \left(  \tA^{p_1}_{x_1 x_2} \, \tB^{q_1}_{x_2x_2} \tA ^{p_2}_{x_2x_3}\cdots \tB^{q_{s-1}}_{x_sx_s}
\tA ^{p_s}_{x_s x_1} 
\tB ^{q_s}_{x_1x_1} \right)
$$
$$
= 
{1\over N} \sum_{x_1=-n }^n \cdots \sum_{x_s=-n }^n  \ \prod_{i=1}^s 
\ \sum_{x_i^{(2)}, \dots, x_i^{(p_i)}}\ 
\ \prod_{j=1}^s\
 \sum_{y_j^{(1)}, \dots, y_j^{(q_j)}} 
 \left({v\over \sqrt {\phi_1}}\right)^{P+2Q}
$$
$$
\times
\E \left( a_{x_i x_i^{(2)}}\,
 a_{x_i^{(2)} x_i^{(3)}}\, \cdots \,
 a_{x_i^{(p_{i})} x_{i+1}†} \cdot 
 a_{x_{j+1}y_j^{(1)}} \,
a_{x_{j+1}y_j^{(2)}}\,
\cdots\, 
a_{x_{j+1}y_j^{(q_j)}}\right).
\eqno(2.8)
$$

Regarding the product of random variables of (2.8), we cannot 
 say do they represent the  independent ones or not;
 therefore we cannot compute the mathematical expectation
 until  
 concrete values taken by variables $x$ and $y$ are known.
Given these concrete  values that we denote by $\langle x\rangle$ and $\langle y\rangle$,
we can use the fundamental property that 
 the random variables
 $a_{\langle x \rangle \langle y\rangle }$ and 
 $a_{\langle  x'\rangle  \langle y'\rangle }$ (1.5) 
 are independent 
 and all  their mixed moments factorize 
 unless either $\langle  x\rangle =\langle  x'\rangle $ and 
 $\langle  y\rangle = \langle y'\rangle $ or $\langle  x\rangle =
 \langle y'\rangle $ and $\langle y\rangle =\langle x'\rangle $.
We are going to  develop a diagram technique that  helps to compute the mathematical expectation 
of (2.8) for any given particular realization of variables 
$\bar x_{\bar p}^{(s)}$ and $\bar y_{\bar q}^{(s)}$.
Using these diagrams, we can  separate the set of all such 
realizations  into the classes of equivalence
 and to evaluate  the sum over
those classes that give non-vanishing contribution to the sum (2.8)
in the limit $(N,R)^\circ\to\infty$. 

The diagram technique we develop is not completely new. We are based on the approach proposed in papers 
\cite{KSV,KV} to study the moments of the discrete version of the Laplace operator for large random Erd\H os-R\'enyi graphs.
However, the fact that the edge probability of $e=\{x,y\}$ depends 
on the difference $x-y$ (1.5) requires a number of essential modifications
of the method of \cite{KSV,KV}.
The main idea here is to describe a kind of multigraph   such that its edges
represent  independent random variables and such that the multiplicities of the multi-edges
take into account the number of times that given random variable 
appears in the product  (2.8). We refer to these multigraphs as to the diagrams. Their structure is fairly natural and intuitively clear
(see examples presented below on Figures 1 and 2), while the
formal description of their construction  is somehow  cumbersome. 
Let us pass to rigorous definitions.

\subsubsection{Color diagrams}
Given $\bar p$, $\bar q$ and  regarding the family of variables 
$(X_{\bar p},Y_{\bar q})=(\bar x_{\bar p}^{(s)},
 \bar y_{\bar q}^{(s)})$ of the right-hand side of (2.8), we denote by 
 $$
 \langle X_{\bar p},
 Y_{\bar  q}\rangle_n
= \langle \bar x_{\bar p}^{(s)}, \bar y_{\bar q}^{(s)}\rangle_n
$$ 
a particular realization   of $(X_{\bar p},Y_{\bar q})$ that attributes to each variable an integer  from  $\{-n,\dots, n\}$
 and construct by recurrence 
a diagram 
$\CD_k(\langle X_{\bar p},
 Y_{\bar  q}\rangle_n)$  that is a multigraph 
with $k$ oriented edges that we denote by $e_i, i=1, \dots,  k$.
To simplify denotations, we will omit the subscripts and arguments
in $\CD$
when no confusion can arise.
We denote the set of edges of $\CD$ by $\CE$ and the set of
 vertices of $\CD$ by  $\CV=\{\u_j, 1\le j\le \kappa\}$.
The diagram $\CD = (\CV, \CE)$ consists of two sub-diagrams: 
the blue one $\CD^{(b)}$
with blue vertices  and $P$ blue edges,
 and the red one $\CD^{(r)}$
with red vertices  and $Q$ red edges.

\vskip 0.2cm 
We construct $\CD^{(b)}$ first. To do this, we consider 
the realization $\langle X_{\bar p}\rangle_n$ 
as an ordered sequence of $P+1$ integers denoted by 
$\CI_{P+1}= (I_1,I_2, \dots, I_{P}, I_1)$, $I_i\in \{-n,\dots, n\}$.
This order is naturally imposed by variables $x_i^{(j)}$ in the product
of (2.6).
We start the construction of $\CD^{(b)}$
by drawing  the vertex  $u_1$;
we attribute to it the integer $I_1= \langle x_1^{(1)}\rangle_n$.
This is the first step of the recurrence that we will refer to as 
the $x$-recurrence.

If $p_1>0$, then the second integer $I_2$ is given  by 
$\langle x_1^{(2)}\rangle_n$. If  $p_1=0$, then 
 $I_2= \langle x_2^{(2)}\rangle_n$. 
It follows from (1.5) that $a_{xx}^{(R)}=0$;  to avoid the trivial case, we accept that  $I_2\neq I_1$ 
 and  create the second  vertex  
$u_2\neq u_1$. Then we attribute to $u_2$ the integer $I_2$ and 
draw the oriented edge 
$\ep_1 = (u_1,u_2)$. We attribute to this edge the random variable
$a_{I_1I_2}$.
This terminates the second step of the $x$-recurrence, $t=2$. 

To describe the general recurrence rule, we consider the  ensemble  $\bU_t$
 of vertices $u_i$ created during the first $t$ steps and denote 
 $\varkappa_t= \# \bU_t$, $t\ge 1$. We denote by $\bI_t$ the set of integers
 attributed to the vertices of $\bU_t$. We assume that  
 all elements of $\bI_t$ are different and $\bI_t$ is in one-to-one correspondence
 with $\bU_t$. These properties will be  proved by construction during  the $x$-recurrence procedure.  
 
 We take the next in turn integer
 $I_{t+1}$ not equal to $I_t$ due to (1.5) and perform the following actions:

  - if there exists such $I' \in \bI_t$  that $I_{t+1}= I'$,
then we consider 
the already existing vertex $u'=u(I')$ attributed by 
$I'$; this vertex is uniquely determined. Then we consider 
the vertex $\hat u=u(I_t)$ attributed by $ I_t$
and draw
the edge $\ep_{t} = (\hat u, u')$. It is clear that 
in this case  $\varkappa_{t+1}=\varkappa_t$.
We attribute to this edge the random variable $a_{I_t I'}$;

  - if there is no such values $I'\in \bI_t$ that $I'=I_{t+1}$, then we create 
  a new vertex $u_{\varkappa_t+1}$, attribute to it 
  $I_{t+1}$ 
  and draw the edge $\ep_{t}= (\hat u, u_{\varkappa_t+1})$, where 
  $\hat u = u(I_t)$.
 Obviously,  $\varkappa_{t+1}=\varkappa_t+1$.
 We attribute to this edge the random variable $a_{I_t I_{t+1}}$.

We proceed till the last step $t+1=P+1$ is performed 
and the last edge $\ep_P= (u(I_P), u_1)$ is drawn.  We color all 
vertices of $\bU_{P+1}$ and $P$ edges obtained in blue color. 
The sub-diagram  
$\CD^{(b)}$ is constructed. 

\vskip 0.2cm
Now we describe the $y$-recurrence that adds to $\CD^{(b)}$
the red elements of $\CD$. 
Realization $\langle \bar y_{\bar q}^{(s)}\rangle$
can be considered as a chronologically ordered sequence of $Q$ integers $\CJ=(J_1,\dots, J_Q)$.
We take the first integer $J_1 =\langle y_1^{(1)}\rangle_n$ and compare it with the elements of 
$\bI_P$. Taking into account (1.5), we assume that $J_1\neq I_{p_1+1}$.
If there exists such $I'\in \bI_P$ that 
$J_1 = I'$, then we join the vertex $\hat u\in \bU_P$ attributed by 
$I_{p_1+1}$,  with the vertex $u'$ attributed by
$I'$ by the edge $(\hat u,u')$ and color it in red.
We attribute to this edge random variable $a_{I_{p_1+1}I'}$.
If there is no such $I'$, then we create a new  vertex $w_1$
that we color in red and draw the red edge
$\vep_1 = (\hat u, w_1)$. We attribute to $w_1$ the value $J_1$ and 
attribute to $\vep_1$ the random variable $a_{I_{p_1+1}J_1}$.

To describe  the general step $\t+1$ of $y$-recurrence, 
we introduce the set $\bJ_\tau$ of integers attributed to the 
set $\bW_\tau$ of red vertices created during the first 
$\tau$ steps. We take $J_{\tau+1}$ and compare it
with the elements of the set $\{\bI_P, \bJ_\tau\} = \bI_P\cup \bJ_\tau$.

- If there exists such $K'\in \{\bI_P, \bJ_\tau\} $ that 
$J_{\tau+1}=K'$,  then we draw the red edge 
$\vep_{\tau+1} = (\hat u, v')$, where 
$\hat u = u(I_{p_1+\dots+p_j+1})$ if $q_1+\dots+q_{j-1}+1\le \tau
\le q_1+ \dots+q_{j}$ and
$$
v' = 
 \begin{cases}
  u'=u(K') \in \bU_P , & \text{if  $K'\in \bI_P$} , \\
w'=w(K')\in \bW_\tau, & \text {if $K'\in \bJ_\tau$.}
\end{cases} 
$$
 In this case the cardinality $\varrho_\tau = \#\bW_\tau$
 does not change its value, $\varrho_{\tau+1} = \varrho_\tau$.
 We attribute to the edge $\vep_{\t+1}$ 
 the random variable $a_{I_{p_1+\dots +p_j+1}K'}$.
  
  - If there is no such $K'\in \{\bI_P, \bJ_\tau\} $ that 
$J_{\tau+1}=K'$,  then we create a new red vertex 
$w_{\varrho_\tau +1}$ and attribute $J_{\tau+1}$ to it.
We also draw the red edge
$\vep_{\tau+1} = (\hat u, w_{\varrho_\tau+1})$
 and set $\bJ_{\tau+1} = \{ \bJ_\tau, J_{\tau+1}\}$.   We attribute to the edge $\vep_{\t+1}$ 
 the random variable $a_{I_{p_1+\dots +p_j+1}J_{\t+1}}$.
 In all these considerations, we have assumed that $J_{\t+1}\neq I_{p_1+\dots +p_j+1}$.
 
  The last step of the $y$-recurrence is $\tau+1=Q$ and 
  the red sub-diagram $\CD^{(r)}$ is constructed.
  Let us point out that the total number of vertices 
  $\nu=\#\CV$ cannot be greater than $n$.
   
  Finally, we get
  the diagram $\CD=(\CV,\CE)$ by renaming 
  the edges  
  of  $\CD^{(b)}\uplus \CD^{(r)}$ by $(e_1,e_2,\dots , e_{k})$,
  $k=P+Q$ according to the data $(\bar p,\bar q)$. 
  This means that the first $p_1$ edges $e_1, \dots, e_{p_1}$ are the blue ones, then $e_{p_1+1},\dots, e_{p_1+q_1}$ are the red ones followed by 
  $p_2$ blue edges $e_{p_1+q_1+1}, \dots, e_{p_1+q_1+p_2}$, 
  and so on.  Finally, we rename the vertices 
  of $\CD^{(b)}\uplus \CD^{(r)}$ by $(\u_1, \u_2, \dots, \u_\nu)$
  according to the order of their appearance in the sequence
  of edges $(e_1,e_2,\dots , e_{k})$. The diagram $\CD$ is constructed.

 \begin{figure}[htbp]
 \centerline{\includegraphics[width=12cm,height=8cm]
 {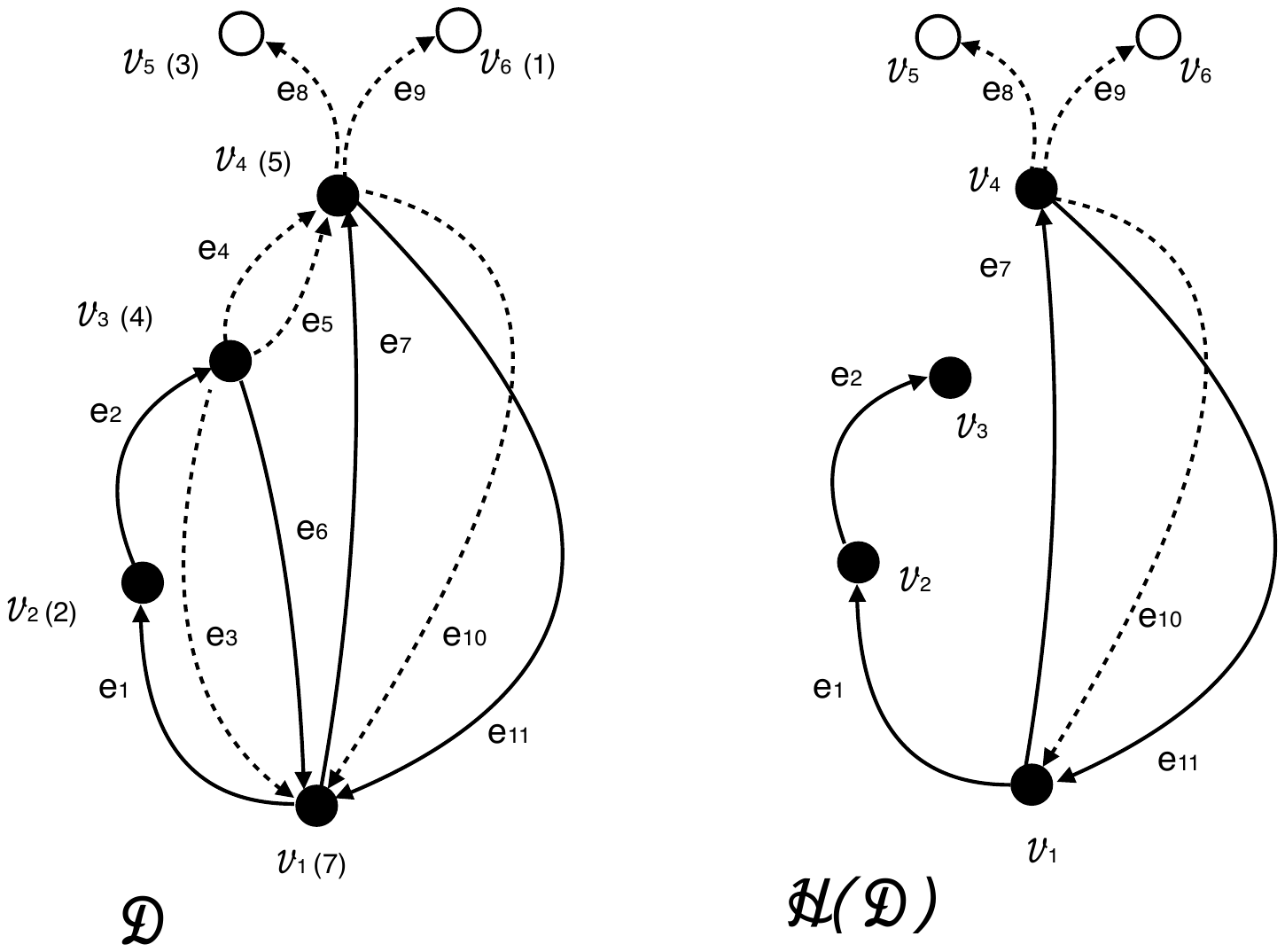}}
 \caption{\footnotesize{ The  diagram $\CD$ and its sub-diagram 
 $\CH(\CD)=\hat \CD$}}
 \end{figure}

\vskip 0.2cm

For example, taking $\bar p=(2,2,1)$, $\bar q = (3,3,0)$
and realization $\langle X_{\bar p},Y_{\bar q}\rangle_{9}$  such that 
$$
\CI= (7,2,4,7,5,7), \quad \CJ = (7,5,5,3,1,7),
$$
we get a diagram $\CD$ 
with  six vertices, five blue edges $e_1,e_2,e_6,e_7, e_{11}$ and six red edges $e_3,e_4,e_5,e_8,e_9,e_{10}$,
where $e_1=\epsilon_1=(\u_1,\u_2)$, $e_2=\epsilon_2=(\u_2,\u_3)$, $e_3=\varepsilon_1=(\u_3,\u_1)$, $e_4=\varepsilon_2=(\u_3,\u_4)$ and so on.  
On  Figure 1, we denoted by dotted lines and empty circles
the red edges and  red vertices, respectively.
The numbers in parenthesis indicate the integers attributed to the vertices of $\CD$.
For simplicity, we have used in $\CI$ and $\CJ$ positive integers only. 

\vskip 0.2cm

As one can see, the diagram $\CD$  contains $P$ blue edges and $Q$ red ones.  For any two vertices $\u_i $ and $\u_j$  joined by an edge of $\CD$,
we denote by 
$\fb(\{\u_i, \u_j\})$ the total number of blue edges of the form $(\u_i,\u_j)$ and $(\u_j, \u_i)$ and by $\fr(\{\u_i, \u_j\})$ the total number of red edges of the form $(\u_i,\u_j)$ and $(\u_j, \u_i)$.

Regarding a vertex $\u\in \CV(\CD)$ such that  
$\u\neq \u_1$, one can find  unique vertex $\u'$ such that 
$\u$ has been  created in $\CD^{(b)}\uplus \CD^{(r)}$
by arrival from $\u'$. Keeping all edges of $\CD$ of the form
$(\u,\u')$ and $(\u',\u)$ and erasing all other edges from 
$\CE(\CD)$, we will get a new diagram that is a sub-diagram of $\CD$.
 We denote this sub-diagram by 
$\hat \CD = \CH(\CD)$ (see Figure 1).

Regarding the set of vertices   $\CV(\CD)$, we can join 
each couple $\u_i,\u_j$ such that $\fb(\{\u_i, \u_j\})+\fr(\{\u_i, \u_j\})>0$ by a non-oriented simple edge.  
Thus we get a graph $G= G(\CD)$
 such that  $\CV(G)=\CV(\CD)$. We accept that 
 the vertices of $G$ are ordered according to the order of elements
of $\CV(\CD)$ and denote the set of ordered edges of $G$ by $E(G)$.
Regarding $G(\hat\CD)$, we get a graph $\hat G$
 that is a plane rooted tree with the root vertex $\u_1$.

 Having constructed the diagram
$\CD$, 
we define a generalized diagram $ \CD_\xi$ 
by attributing variables $\xi_i, 1\le i\le \nu$ to the vertices
of $\CD$
according to their order.
In what follows, we denote the collection of these variables  by 
 $\bar \xi_\nu=(\xi_1,\dots,\xi_\nu)$. By the same way 
 we determine the generalized graph $G_\xi$.

\subsubsection{Weight of the diagram}

By construction,  random variables $a$ (1.5) attributed to different 
edges of $G(\CD) $ are different and therefore they  are 
 jointly independent. 
Returning to the  example realization $\langle X_{\bar p},Y_{\bar q}\rangle_{9}$  from above and considering the corresponding 
diagram $\CD$ of Figure 1, 
 we can easily find that the value of the 
average of the right-hand side of (2.8) 
is equal to the mathematical expectation 
$$
{v^{11}\over \phi_1^{11/2}}\E\left(
a^{(R)}_{72}a^{(R)}_{24} \left( a^{(R)}_{47}\right)^2
\left( a^{(R)}_{45}\right)^2 
 a^{(R)}_{53} a^{(R)}_{51}
\left( a^{(R)}_{57}\right)^3\right)
$$
that factorizes  to the product of  moments of seven random variables involved.

Regarding the generalized diagram 
$\CD_\xi$, it is natural to attribute to each non-oriented 
edge $e= \{\u_i,\u_j\}$ of  $G(\CD_\xi)=G_\xi$ 
the value 
$$
\pi_e^{(R)}(\bar \xi_\nu) = \pi^{(R)}_{\{\u_i,\u_j\}}(\bar \xi_\nu) = 
 {\displaystyle v^{\fb(e)+2\fr(e)}\over\displaystyle   \phi_1^{\fb(e)/2+\fr(e)}   }\cdot {1\over R} \phi( (\xi_i- \xi_j)/R)
 \eqno (2.9)
$$
where  $\xi_i$ and $\xi_j$ are given by the  variables attributed to 
the vertices $\u_i$ and $\u_j$, respectively. 
Then the total weight of the generalized diagram is 
$$
\Pi^{(R)}(\CD_\xi, \bar \xi) = 
\prod_{e\in G_\xi}  \pi^{(R)}_{\{\u_i,\u_j\}}(\bar \xi_\nu).
\eqno (2.10)
$$

Given $\bar p$ and $\bar q$ of (2.8), any realization of variables 
$\langle X_{\bar p},  Y_{\bar q}\rangle_n$
generates a diagram $\CD(\langle X_{\bar p},  Y_{\bar q}\rangle_n)$. 
We say that two realizations 
$\langle X_{\bar p},  Y_{\bar q}\rangle_n'$ and 
$\langle X_{\bar p},  Y_{\bar q}\rangle_n''$ are equivalent, if their diagrams coincide. 
One should note that the value of the mathematical expectation (2.8)
for these realizations can be different. From another hand, 
for any realization $\langle X_{\bar p},  Y_{\bar q}\rangle_n$
a  unique realization $\langle \bar \xi_\nu\rangle_n$
of the variables of $\CD_\xi$ is determined; obviously, this realization is
given by the values attributed to  the vertices of $\CD(\langle X_{\bar p},  Y_{\bar q}\rangle_n)$. Inversely, any pair $(\CD_\xi, \langle 
\bar \xi_\nu\rangle_n)$ determines uniquely a realization 
$\langle X,Y\rangle_n$. 

Let us denote by $\CC(\CD)$ the class of equivalence of realizations
corresponding to $\CD$. Clearly, $\#\CC(\CD) = N(N-1)\cdots (N-\nu+1)$,
where $\nu = \# \CV(\CD)$.
The contribution of the class 
$\CC(\CD)$ to the sum (2.8) is given by expression
$$
 \Pi^{(N,R)}(\CD,  \Xi_\CD^{(N)})  = 
 {1\over N} \sum_{\xi_1: \vert \xi_1\vert\le n}\ 
  \Pi^{(N,R)}(\CD,  \Xi_\CD^{(N)}(\xi_1)),
  \quad N=2n+1,
   \eqno(2.11)
$$
 where 
 $\Xi_\CD^{(N)}$ is the set of all possible realizations 
 $\langle\bar \xi_\nu\rangle_n$ such that 
$\langle \xi_i\rangle_n\neq\langle \xi_j\rangle_n$
for any $i,j$ such that $\{\u_i, \u_j\}\in E(G(\CD))$,
 $$
  \Pi^{(N,R)}(\CD,  \Xi_\CD^{(N)}(\xi_1))= 
  \sum_{
 \langle\bar \xi_{\nu}\rangle_n^{(1)}\in \Xi_\CD^{(N)}(\xi_1)} 
 \Pi^{(R)}( \CD_\xi,\langle \bar \xi_\nu\rangle_n^{(1)}),
 $$
$\langle \bar \xi_\nu\rangle_n^{(1)} = \langle \xi_2, \dots, \xi_\nu\rangle_n$ 
and  
$$
\Xi_\CD^{(N)}(\xi_1)=\left\{ \langle \xi_2, \dots, \xi_\nu\rangle_n:  \langle \xi_i\rangle_n\neq\langle \xi_j\rangle_n {\hbox{ for $i,j$ such that }} \{\u_i, \u_j\}\in E(G(\CD))\right\}.
$$
Then we can rewrite (2.6) and (2.8)  in the form
$$
 \CS^{(N,R)}_{(\bar p, {\bar q})}(s)
=\sum_{\CD\in \bD_{(\bar p, \bar q)}}
\Pi^{(N,R)}(\CD, \Xi_\CD^{(N)}),
$$
where $\bD_{(\bar p, \bar q)}$ denotes the family of all
 diagrams with given  $\bar p$ and $\bar q$.
It follows from (2.7) that 
$$
M_k^{(N,R)} = \sum_{s=1}^k\  \sum_{(\bar p,\bar q)^*}^k
\  \sum_{\CD\in \bD_{(\bar p, \bar q)}} (-1)^P\,
\Pi^{(N,R)}(\CD, \Xi_\CD^{(N)}),
\eqno (2.12)
$$
where the sum over $(\bar p, \bar q)^*$ runs over the set determined  
in (2.6).

\subsection{Contributions of diagrams and tree-type walks}

In this subsection we study $\Pi^{(N,R)}(\CD, \Xi_\CD^{(N)})$
 (2.11) 
in the limit of infinite $N$ and $R$ (1.10).
We are going to prove that  non-vanishing 
contribution to (2.12) is given by 
such  diagrams 
$\CD$ whose graphs $G(\CD)$ have no cycles.
In denotations of previous subsection, this means that 
$\CD=\hat \CD$, where $\hat \CD= \CH(\CD)$. In this case $G(\CD)=G(\hat \CD)$
 and
 we will say that 
the diagram $\CD$  is of the tree-type structure. 

\vskip 0.3cm

{\bf Lemma 2.1.} {\it If $\CD$ is such that $G(\CD)$ has a cycle, $\CD\neq \hat\CD$, then 
$$
\Pi^{(N,R)}(\CD, \Xi_\CD^{(N)}) = 
o\left(\Pi^{(N,R)}(\hat \CD, \Xi_\CD^{(N)})\right), \quad (N,R)^\circ\to\infty,
\eqno (2.13)
$$
where 
$$
\Pi^{(N,R)}(\hat \CD, \Xi_\CD^{(N)}) =  
\prod_{e\in G(\hat \CD)} {v^{\fb(e)+2\fr(e)}\over \phi_1^{\fb(e)/2+\fr(e)-1}} 
 (1+ o(1)), \quad (N,R)^\circ\to\infty.
\eqno (2.14)
$$
If $G(\CD)$ is a tree, then each blue edge  $e$ of $G(\CD^{(b)})$   has  an even multiplicity $\fb(e)=2\ell(e)$ and  (2.14) is valid with 
$\hat \CD$ replaced by $\CD$,
$$
\Pi^{(N,R)}( \CD, \Xi_\CD^{(N)}) =
\Pi(\CD) (1+ o(1)), \quad (N,R)^\circ\to\infty,
\eqno (2.15),
$$
where we have denoted
$$
\Pi(\CD)=   
\prod_{e\in G(\CD)} {v^{2(\ell(e)+\fr(e))}\over \phi_1^{\ell(e)+\fr(e)-1}}.
$$

}

\vskip 0.3cm

{\it Proof of Lemma 2.1.} 

 We start with the proof of (2.13). 
 Given $\CD$, let us  consider an edge $e= \{\gamma, \delta\}$ of $G(\CD)$ that 
does not belong to $E(\hat G)$. 
Taking into account that $0<\phi(t)<1$, 
we  estimate the weight of this edge by a trivial  inequality 
$$
 \pi^{(R)}_{\{\g,\d\}}(\bar \xi) \le 
 {\displaystyle v^{\fb(e)+2\fr(e)}\over\displaystyle   
 R \, \phi_1^{\fb(e)/2+\fr(e)}   }.
$$
Then clearly,
$$
\Pi^{(R)}(\CD_\xi, \bar \xi_\nu) \le {v^r \over \phi_1^{r/2} \, R^{\vert E(G)\vert - \vert E(\hat G)\vert} }
 \Pi^{(R)}( \hat \CD_\xi, \bar \xi_\nu)
$$
with some $r>0$. Summing the last inequality  over all 
realizations of $\bar \xi_\nu$, we get    (2.13).   

\vskip 0.2cm

Let us count the multiplicities of blue edges in tree-type diagram 
$\CD$.
 To do this, 
we consider a sub-diagram $\CD^{(b)}$ of $\CD$ that contains
only blue vertices and blue edges of $\CD$ and determine  
by the same rule as before the graph $G^{(b)} = G(\CD^{(b)})$.
Let us recall that by construction, the 
sequence of blue edges of $\CD$ makes a closed circuit 
that starts and ends at the root vertex $u_1$. 
  
If $G=G(\CD)$ is a tree, $G=\CT$, then $G^{(b)}= G(\CD^{(b)})$ is also a tree,
$G^{(b)}=\CT^{(b)}$. 
Let us prove that in this case 
each edge  $\{u_i, u_j\}\in E(G^{(b)})$ is passed by
the edges of $\CD^{(b)}$ an even number of times
and the number of edges of the form $(u_i,u_j)$ is equal
to the number of edges of the form $(u_j,u_i)$. Let $u_j\neq u_1$ be
a vertex that has only one edge $e= \{u_j, u_l\}$ attached to it.
Such vertex is referred to as the leaf of the tree $\CT^{(b)}$. 
By the construction, there exists at least one blue edge of $\CD$
ending at $u_l$ and  this edge is  necessarily of the form 
$e'=(u_l,u_j)$. 
Let us assume that $e'$ is the last edge of $\CD^{(b)}$ of this form.
Since $u_l\neq u_1$, there should be at least one edge $e''$ of 
$\CD^{(b)}$ of the form $e''=(u_j,u_l)$. Let us assume that this is the first
edge of $\CD^{(b)}$ of this form. Now we can reduce 
the diagram $\CD^{(b)}$ to the diagram 
$\tilde \CD^{(b)}$ by removing 
these edges $e'$ and $e''$. If $e'$ is the unique arrival at $u_j$,
then $V(\tilde \CG^{(b)}) = V(\CG^{(b)})\setminus u_j$, otherwise $u_j$ remains in $V(\tilde \CG^{(b)})$.
 It is easy to see that $G(\tilde \CD^{(b)})$ is again a tree 
$\tilde \CT^{(b)}$ and we can repeat this procedure by recurrence
till there is no vertices remaining excepting the root vertex $\u_1$. 
Thus for each edge $e\in \CT^{(b)}$, there exists a natural  $\ell(e)$ such that $\fb(e)=2\ell(e)$.

\vskip 0.3cm

We prove relations (2.14) and (2.15) in the case $\CD=\hat \CD$ of tree-type diagrams only, thus (2.15)
because the proof of  (2.14) is literally the same.
Let us start with  an auxiliary upper bound
$$
  \Pi^{(N,R)}(\CD,  \Xi_\CD^{(N)}(\xi_1))\le \Pi(\CD)
  \eqno(2.16)
$$
that is true for all $N,R$ and $\xi_1$. 
We prove (2.16) by recurrence with respect to the number of edges in 
the tree $\CT=G(\CD)$. Let us recall that the edges of $\CT=\CT_\nu$ as well as its vertices
$\u_j, 1\le j\le \nu $ are naturally ordered according  to the instant of their creation 
during the construction of the diagram $\CD$. Consider the last edge   
$e'= \{\u_{\zeta},\u_\nu\}$.  It follows from (2.9) and (2.10)  that 
$$
\Pi^{(N, R)} (\CD, \Xi_\CD^{(N)} (\xi_1)) =
\sum_{  \langle\bar \xi_{\nu-1} \rangle_n^{(1)}\in \,\Xi_{\CD\setminus [e']}^{(N)} (\xi_1)}
 \Pi^{(R)}(\CD_\xi\setminus [e'],  \langle \bar \xi_{\nu-1}\rangle_n^{(1)}  ) 
 $$
 $$
 \times
 \ \sum_{\xi_\nu:   \xi_\nu\neq \langle\xi_{\zeta}\rangle}  
{1\over R}\phi\left( {\xi_\nu - \langle\xi_{\zeta}
\rangle 
\over R}\right) 
{v^{2(\ell(e')+\fr(e'))}\over \phi_1^{\ell(e')+\fr(e')}},
\eqno (2.17)
$$
where  $[e']$ denotes the family of all edges of $\CD$ of 
the form $(\u_\zeta,\u_\nu)$ and $(\u_\nu,\u_\zeta)$.
Taking into account positivity  of all terms in the right-hand side of (2.17)
and using elementary inequality 
$$
\sum_{s: \, -m\le s\le m',\,  s\neq 0} \ {1\over R}\,  \phi\left({s\over R}\right)\le \int_{-\infty}^\infty \phi(t) dt = \phi_1
\ {\hbox{ for any natural }}m,m', 
$$
we deduce from (2.17) inequality
$$
\Pi^{(N,R)} (\CD, \Xi_\CD^{(N)} (\xi_1)) \le \Pi(\CD\setminus [e']) \ 
{v^{2(\ell(e')+\fr(e'))}\over \phi_1^{\ell(e')+\fr(e')-1}},
$$
where we assumed that (2.16) holds for the diagram $\CD\setminus [e']$.
It remains to verify the first step of recurrence for the diagram $\CD= \{e'\}$ with $e'=[\u_1,\u_2]$, 
$$
\Pi^{(N,R)}(\{e'\}, \Xi^{(N)}_{\{e'\}}(\xi_1))
=  \ \sum_{\xi_2:   \xi_2\neq \xi_{1}}  
{1\over R}\phi\left( {\xi_2 - \xi_{1}
\over R}\right) 
{v^{2(\ell(e')+\fr(e'))}\over \phi_1^{\ell(e')+\fr(e')}}\le  
{v^{2(\ell(e')+\fr(e'))}\over \phi_1^{\ell(e')+\fr(e')-1}}.
$$
The upper bound (2.16) is proved.

\vskip 0.5cm
It follows from (1.5) that given  $\vep>0$, there exists $T>0$ such that 
$$
{1\over \phi_1} \int _{\vert t\vert >T}\phi(t) dt = \vep.
\eqno (2.18)
$$
Remembering that the graph $G(\CD)=\CT$ is a tree, we introduce a set of realizations  
$$
\Xi^{(T,R)}_{\CT}(\xi_1) = 
\left\{\langle\xi_2,\dots, \xi_\nu\rangle_n: \, 
0<\vert \langle \xi_i\rangle_n  - \langle\xi_j\rangle_n\vert\le TR\  
{\hbox { for }} 
\{\u_i,\u_j\}\in E(\CT)\right\}
$$
and  denote
$$
\Pi^{(R)} (\CD, \Xi^{(T,R)}_{\CT}(\xi_1))  = 
\sum_{ \langle\bar \xi_{\nu}\rangle\in \Xi^{(T,R)}_{\CT}(\xi_1)} \Pi^{(R)}( \CD_\xi,\langle \bar \xi_\nu\rangle).
 $$
On the first stage of the proof of (2.15),
we  show   that asymptotic equality 
$$
\Pi^{(R)} (\CD, \Xi_G^{(T,R)} (\xi_1))  = 
\prod_{e\in G} {v^{2(\ell(e)+\fr(e))}\over \phi_1^{\ell(e)+\fr(e)-1}} 
\left(1+o(1) + C_\nu(\vep)\right), 
\ \ R\to\infty
\eqno (2.19)
$$
holds for any $\xi_1\in [-n+\nu TR, n-\nu TR]$
where $C_\nu(\vep)$ vanishes as $\vep\to 0$.

As above, we prove (2.19) by recurrence and write that (cf. (2.17))
$$
\Pi^{(R)} (\CD, \Xi_G^{(T,R)} (\xi_1)) =
\sum_{  \langle\bar \xi_{\nu-1} \rangle_n^{(1)}\in \,
\Xi_{G\setminus e'}^{(T,R)} (\xi_1)}
 \Pi(\CD_\xi\setminus [e'],  \langle \bar \xi_{\nu-1}\rangle_n^{(1)}  ) 
 $$
 $$
 \times
 \ \sum_{\stackrel{\xi_\nu: \, \vert \xi_\nu  -\langle\xi_{\zeta} \rangle\vert\le TR}{\,  \xi_\nu\neq \langle\xi_{\zeta}\rangle} } 
{1\over R}\phi\left( {\xi_\nu - \langle\xi_{\zeta}
\rangle 
\over R}\right) {v^{2(\ell(e')+\fr(e'))}\over \phi_1^{\ell(e')+\fr(e')}}.
\eqno (2.20)
$$
Adding  to the right-hand side of (2.20) the auxiliary term 
$$
 \sum_{ \langle \bar \xi_{\nu-1}\rangle_n^{(1)} 
 \in \Xi_{G\setminus e'}^{(T,R)}(\xi_1) }
\ \Pi(\CD_\xi\setminus [e'], \langle\bar \xi_{\nu-1} \rangle_n^{(1)})\ 
{v^{2(\ell(e')+\fr(e'))}\over \phi_1^{\ell(e')+\fr(e')-1}}
$$
and subtracting it, we get relation 
$$
\Pi^{(R)} (\CD, \Xi_G^{(T,R)} (\xi_1)) = \sum_{  \langle\bar \xi_{\nu-1}\rangle_n^{(1)} 
\in \Xi_{G\setminus e'}^{(T,R)}(\xi_1) }
 \Pi(\CD_\xi\setminus [e'],  \langle\bar \xi_{\nu-1}\rangle_n^{(1)} )
 $$
 $$
 \times
{v^{2(\ell(e')+\fr(e'))}\over \phi_1^{\ell(e')+\fr(e')-1}}
(1+ \chi(R,T) -\vep), 
\eqno (2.21)
$$
where 
$$
\chi(R,T) = {1\over R\phi_1} \sum_{\stackrel{\xi: \vert \xi\vert\le TR}{\xi\neq 0}}
\phi(\xi/R) - {1\over \phi_1}\int_{-T}^T \phi(t)dt. 
$$
It is clear that $\chi(R,T)= o(1), R\to \infty$ for any given $T>0$.
Assuming that 
 (2.19) holds with $G$ replaced by for $G\setminus e'$,
$$
\Pi_N^{(R)}\left(\CD\setminus [e'], \Xi_{G\setminus e'}^{(T,R)}(\xi_1)\right)
=  \prod_{e\in G(\CD)\setminus e'} 
{v^{2(\ell(e)+\fr(e))}\over \phi_1^{\ell(e)+\fr(e)-1}}
 \left(1+o(1)+ C_{\nu-1}( \vep)\right),
$$
and substituting this asymptotic equality into the right-hand side 
of (2.21), we conclude after simple computations that (2.19) is true 
and that $C_\nu(\vep)$ is such that 
$$
C_\nu(\vep) = (1-\vep) C_{\nu-1}(\vep) -\vep.
\eqno (2.22)
$$

On the first step of the recurrence
we get equality
$$
{1\over R}\  \sum_{\stackrel{\xi_2: \vert\xi_1- \xi_2\vert\le RT}{\xi_2\neq\xi_1}}
\phi\left( {\xi_1-\xi_2\over R}\right)
{v^{2(\ell(e_1)+\fr(e_1))}\over \phi_1^{\ell(e_1)+\fr(e_1)}}= {v^{2(\ell(e_1)+\fr(e_1))}\over \phi_1^{\ell(e_1)+\fr(e_1)-1}}
(1+\chi(R,T)   -  \vep). 
$$
We see that   
$C_1(\vep)= -\vep$. Using this fact and assuming that $0<\vep<1/2$, 
we can deduce from  (2.22) inequality 
 $\vert C_\nu(\vep)\vert \le 2\vep$.
 Relation  (2.19) is proved. 

\vskip 0.2cm

To complete the proof of (2.15),
we  write that 
$$
\Pi^{(N,R)}(\CD, \Xi_\CD^{(N)} )
=
\CR_1+\CR_2+\CR_3 
,
\eqno (2.23)
$$
where 
$$
\CR_1=
{1\over N}\  \sum_{\vert \xi_1\vert\le n- \nu TR}\ \  
\sum_{\langle \bar \xi_\nu^{(1)}\rangle_n \in \Xi_G^{(T,R)}(\xi_1)  } 
\Pi ^{(R)}(\CD_\xi, \langle \bar \xi_\nu\rangle_n^{(1)}),
$$
$$
\CR_2=
{1\over N}\  \sum_{\vert \xi_1\vert> n- \nu TR, \, 
\vert \xi_1\vert \le n}\ \  
\sum_{\langle \bar \xi_\nu\rangle_n^{(1)} \in \Xi_G^{(T,R)}(\xi_1)  } \Pi ^{(R)}(\CD_\xi, \langle \bar \xi_\nu\rangle_n^{(1)}),
$$
$$
\CR_3={1\over N}\  \sum_{\vert \xi_1\vert\le n}\ \  
\sum_{\langle \bar \xi_\nu\rangle_n^{(1)} \in\,    
\Xi_\CD^{(N)}(\xi_1)\setminus {\Xi_G^{(T,R)}(\xi_1)}\  } 
\Pi ^{(R)}(\CD_\xi, \langle \bar \xi_\nu\rangle_n^{(1)}).
$$
It follows from (2.19) that 
$$
\CR_1= 
\prod_{e\in G} {v^{2(\ell(e)+\fr(e))}\over \phi_1^{\ell(e)+\fr(e)-1}} 
\left(1+o(1) + C_\nu(\vep)\right), 
\ \ R\to\infty.
\eqno(2.24)
$$
Taking into account inequality (2.16), we conclude that 
$$
\CR_2\le {1\over N}\  \sum_{\vert \xi_1\vert> n- \nu TR, \, 
\vert \xi_1\vert \le n}\ \   \Pi(\CD) \le {2\nu TR\over N} \Pi(\CD).
\eqno(2.25)
$$
To estimate $\CR_3$, let us note that if 
$\langle \xi_2, \dots, \xi_\nu\rangle_n \in \Xi_\CD^{(N)}(\xi_1)\setminus {\Xi_G^{(T,R)}(\xi_1)}$, then for any given $\langle \xi_1\rangle_n$ 
there exists at least one 
edge $\tilde e = \{\u_i,\u_j\}$  of the tree $\CT=G$ such that $\vert \langle\xi_i\rangle - \langle \xi_j\rangle \vert >TR$.
We consider  $\nu-1$ ordered edges of $\CT$, $e_1, \dots, e_{\nu-1}$
and admit that they are oriented in the way that the tail of the edge  $\alpha(e_l)$
is closer to the  root $\varrho$ than the head $\beta(e_l)$. 
 Let us  denote by $\CT'_l$ the sub-tree of $\CT $ such that  $\b(e_l)$
 is the root vertex of $\CT'_l$. Then $\varrho$ does not belong to the set of vertices of $\CT'_l$.
We also determine the complementary  tree  $\CT''_l= \CT \setminus (e_l \uplus \CT'_l)$.  
 
For any $-n\le \xi_1\le n$, we consider the set of realizations 
$$
\tilde \Xi_\CD^{(N,R,T)}(e_l, \xi_1)
= \left\{ \langle \bar \xi_\nu\rangle_n^{(1)} \in \Xi_\CD^{(N)}(\xi_1) , \  
\vert \langle \xi_{\a(e_l)}\rangle_n -
\langle\xi_{\beta(e_l)}\rangle_n\vert >RT
\right\}.
$$
If $e_l$ is such that $\a(e_l)= \varrho$, then we replace 
$\langle \xi_{\a(e_l)}\rangle_n$ by $\xi_1$.
It is easy to see that 
$$
\Xi_\CD^{(N)}(\xi_1) \subseteq \cup_{l=1}^{\nu-1}\  
\tilde \Xi_\CD^{(N,R,T)}(e_l, \xi_1)
$$
and therefore
$$
\CR_3\le \sum_{l=1}^{\nu-1}\  \sum_{\langle \bar \xi_\nu\rangle_n^{(1)} \in 
\tilde \Xi_\CD^{(N,R,T)}(e_l, \xi_1)} \Pi^{(R)}(\CD_\xi; \langle \bar \xi_\nu\rangle_n^{(1)}).
\eqno(2.26)
$$
Now we can write that
$$
\sum_{\langle \bar \xi_\nu\rangle_n^{(1)} \in 
\tilde \Xi_\CD^{(N,R,T)}(e_l, \xi_1)} \Pi^{(R)}(\CD_\xi; \langle \bar \xi_\nu\rangle_n^{(1)})
= \sum_{\vert \xi_{\a(e_l)}-\xi_{\b(e_l)} \vert >RT} {1\over R} \phi\left({\xi_{\a(e_l)}-\xi_{\b(e_l)}\over R}\right)
$$
$$\times\, 
{v^{2(\ell(e_l) + \fr(e_l))}\over \phi_1^{\ell(e_l)+\fr(e_l)}}\,
\Pi^{(R)}(\CD'_l, \Xi_{\CT'_l}^{(N)}(\xi_{\a(e_l)}))
\, \Pi^{(R)}(\CD''_l, \Xi_{\CT''_l}^{(N)}(\xi_{\b(e_l)})),
\eqno(2.27)
$$
where $\CD'_l$ and $\CD''_l$ denote the sub-diagrams of $\CD$ that correspond to 
$\CT'_l$ and $\CT''_l$, 
respectively.
The last two factors of (2.27) can be estimated from above
with the help of (2.16). Then 
we deduce from (2.26) the following upper bound,
$$
\CR_3 \le 2\int_{T-1}^\infty \phi(t)dt\  \sum_{l=1}^{\nu-1}  
{v^{2(\ell(e_l) + \fr(e_l))}\over \phi_1^{\ell(e_l)+\fr(e_l)}}
\, \prod_{e'\in G(\CD')} {v^{2(\ell(e') + \fr(e'))}\over 
\phi_1^{\ell(e')+\fr(e')-1}}\,
$$
$$
\times \prod_{e''\in G(\CD'')} {v^{2(\ell(e'') + \fr(e''))}\over 
\phi_1^{\ell(e'')+\fr(e'')-1}}.
$$
Denoting ${2\over \phi_1} \int_{T-1}^\infty \phi(t)dt= \tilde \vep$, we get inequality
$$
\CR_3 \le \tilde \vep (\nu-1) \Pi(\CD).
\eqno (2.28) 
$$
Since $\vep$ and $\tilde \vep$ can be made arbitrarily small by the choice of
sufficiently large $T$, we deduce from (2.23), (2.24), (2.25) and (2.28) 
relation (2.15). Lemma 2.1 is proved. $\Box$

\subsection{Total weight of tree-type closed walks}

Combining (2.12) with (2.15),
we can write that 
$$
M_k^{(N,R)} = 
\sum_{s=1}^k\  \sum_{(\bar p,\bar q)^*}^k 
\ \sum_{\CD \in \hat \bD_{(\bar p,\bar q)}}
(-1)^P \Pi(\CD) ( 1+ o(1)) + \CQ_N^{(R)}, \ (N,R)^\circ\to \infty ,
\eqno (2.29)
$$
where $\hat \bD_{(\bar p, \bar q)}$ denotes the set of all
tree-type diagrams with given $\bar p$ and $\bar q$
and 
$$
\CQ_N^{(R)} = 
\sum_{s=1}^k\  \sum_{(\bar p,\bar q)^*}^k 
\ \sum_{\CD \in \bD_{(\bar p, \bar q)}\setminus  \hat \bD_{(\bar p,\bar q)}}
(-1)^P \Pi(\CD, \Xi_D^{(N)}).
$$
Using (2.13), (2.14) and (2.16) and taking into account that the cardinality
$\#\bD_{(\bar p, \bar q)}$ is finite, we conclude that 
$\CQ_N^{(R)}= o(1), (N,R)^\circ \to\infty$.

Remembering that in the tree-type diagrams the multiplicities of 
edges are such that 
$$
\sum_{e\in G(\CD)} \left( 2 \ell (e) + \fr(e)\right) = P+Q=k
$$
and  $P$ is even, $P=2\ell$, we can rewrite (2.29) in the form 
$$
M^{(N,R)}_k = \sum_{\CD \in \hat \bD_k} \Pi (\CD) + o(1),\  (N,R)^\circ\to\infty,
\eqno (2.30)
$$
where $ \hat \bD_k$ denotes the family of all tree-type diagrams that 
have in total $k$ edges.

 In this subsection 
we compute the leading term of the right-hand side of  (2.30).
We consider the tree-type diagrams
only and omit the hats in denotations $\hat \CD$ and $\hat \bD_k$
when they are not crucial.


 \begin{figure}[htbp]
 \centerline{\includegraphics[width=12cm,height=8cm]
 {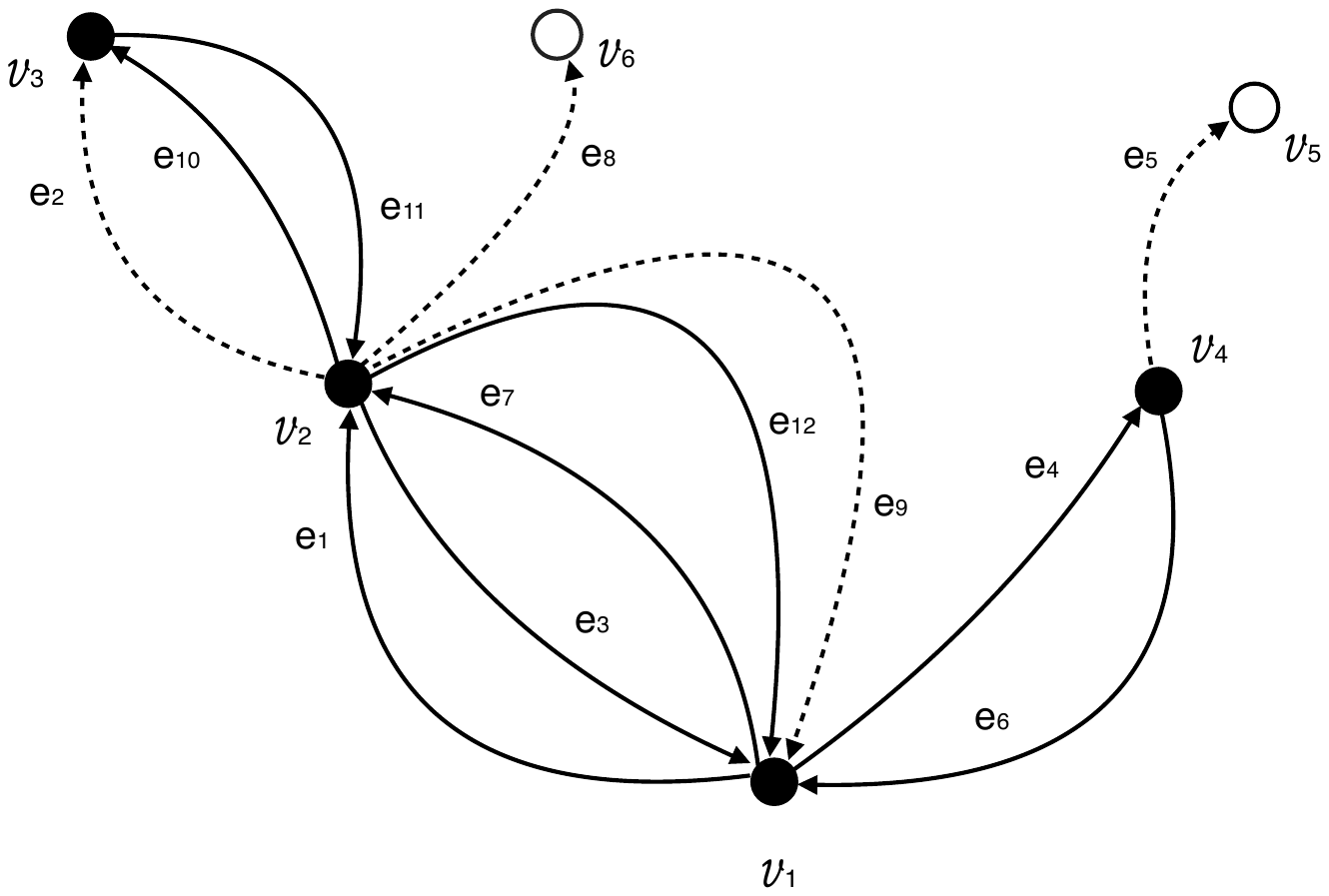}}
 \caption{\footnotesize{The tree-type diagram $\CD=\hat \CD$}}
 \end{figure}

 \subsubsection{Diagrams and close walks}
 
 Given a  diagram $\CD$, one can obtain 
a sequence of $k+1$ letters from the ordered alphabet $\bU= \{\u_1, \u_2, \dots\}$
 by starting at the root vertex $\u_1$ and recording the vertices 
 that appear during the chronological run over $\CD$ determined by its ordered edges. 
 We denote this sequence of letters by $\CW=\CW(\CD)$ and say that 
 each edge $e_t$, $1\le t\le k$ of $\CD$
represents the $t$-th step
 of the walk $\CW$.
 The only convenience is that in the walk constructed 
 we add an additional inverse step $(\u'', \u')$ that follows 
 immediately after that the direct step $(\u',\u'')$ is performed
 along the red edge of $\CD$. 
 One can 
 say that the inverse step $(\u'',\u')$ 
 is the mute one and denote this couple of steps
  by
 $[\u'\u''\u']$.  We consider $[\u'\u''\u']$ as one generalized step
 and say that the vertex $\u''$ is represented in $\CW$ by a generalized letter 
 that we denote by brackets $[\u'']$.
 Thus the total number of ordinary and  generalized steps
  is equal to the number of edges in $\CD$, $P+Q=k$
 and  the walk $\CW$
 is represented as a sequence of  $k+1$ generalized letters.
 It is natural to say that the walk is closed and that it performs  either  blue step 
 or red step when it goes along the blue edge (once) and 
 along the red one (twice), respectively.
 Regarding the diagram $\CD=\hat \CD$ of Figure 2,
we get the closed walk
$$
\CW_{12} (\hat \CD)= 
(\u_1,\u_2, [\u_3], \u_1, \u_4, [\u_5], 
\u_1, \u_2, [\u_6],[ \u_1], \u_3, \u_2, \u_1).
\eqno (2.31)
$$

Given a  diagram $\CD$,
the walk $\CW(\CD)$ is uniquely determined.
If $\CD$ is a tree-type diagram, we say that $\CW(\CD)$  is a tree-type walk. 
We determine the set of all possible tree-type walks by equality
$$
\hat \bW_k = \left\{\CW: \CW(\CD), \CD\in \hat \bD_k\right\}.
$$
 We attribute  to a walk $\CW$ its 
weight  determined by relation
 $\Pi(\CW)=\Pi(\CD)$, $\CW=\CW(\CD)$ (see (2.15)).

Inversely, one can  define a closed walk $\CW_k=(\u(1),  \dots, \u(k+1))$ 
 as a sequence of letters taken from the ordered alphabet $\CU=\{\u_1, \u_2, \dots\}$
 that starts and ends with $\u_1$, 
 $$
 \u(1)= \u(k+1)=\u_1,
 $$
 verifies condition $\u(t+1)\neq \u(t)$ for all $t$, $1\le t\le k$;
we assume that $\CW_k$ is  such that at each instant of time $t+1$ the value $\u(t+1)$
either belongs to the set $\bU_t$ of already existing letters
or is equal to a new letter $\u_{1+ \# \bU_t}$. Some letters
of the collection $\{\u(t), 2\le t\le k\}$
can be  marked as the special (or generalized) ones that we put into the brackets. The remaining non-marked letters are referred to as the ordinary ones.

Regarding a walk $\CW_k$, we can uniquely determine $\CD_k= \CD(\CW_k)$. To do this, we perform the chronological run along $\CW_k$ and draw the vertices and edges between them by the rules described in sub-section 2.1  when constructing $\CD$ for a realization of $k+1$ integers. 
If   $\u(t)$ is marked as the generalized letter, we 
find the maximal value $t-j$, $j\ge 1$ such that $\u(t-j)$ is ordinary,
 draw one direct edge 
$(\u(t-j), \u(t))$ 
and return immediately to the vertex $\u(t-j)$ by the mute step
 without drawing the inverse edge $(\u(t),\u(t-j))$. 
In the diagram obtained, we color the edges that lead to the generalized letters  in red, the remaining edges are colored in blue.

If the diagram $\CD(\CW)$ is of the tree-type, we say that $\CW$ is 
also the tree-type walk. We denote the set of all tree-type walks
of $k$ steps
by 
$\hat \bW_k$. Now we can say that the set of all tree-type diagrams $\hat \bD_k$
and the set of all tree-type walks $\hat \bW_k$ are in one-to-one correspondence.
The same concerns the families $\hat \bD_k$ and $\hat \bW_k$. 
 Then equality 
 $$
 \sum_{\CD\in \hat \bD_k} \Pi(\CD) = \sum_{\CW\in \hat \bW_k} \Pi(\CW)= \Pi(\hat \bW_k)
 $$
together with  (2.30) implies relation
 $$
 M_k^{(N,R)} = \Pi(\hat \bW_k) + o(1),\  (N,R)^\circ\to\infty.
 \eqno (2.32)
 $$

\subsubsection{Total weight  of tree-type walks}

To describe the family of all possible tree-type walks $\hat \bW_k$ and to compute the total weight $\Pi(\hat \bW_k)$, we develop a kind of recurrence
 based on the so-called first-edge decomposition procedure.
 The basic idea
 of this technique dates back to the pioneering paper \cite{W}. In our context, we follow in major part
 the method developed for the tree-type walks in \cite{KSV,KV}.

Let us consider   the root vertex  $\u_1$ of the diagram $ \CD(\CW)$  and the first edge $\{\u_1,\u_2\}$ created by the  first step of any walk $\CW$
of $k$ steps.
If $k=0$, then $\u_2=\u_1$, if $k\ge 1$, then $\u_2\neq \u_1$.
To briefly describe the essence of the first-edge recurrence method,
let us  consider two sub-walks of $\CW$ that we denote by $\dot \CW$ and
$\ddot \CW$. The first one produces 
a diagram $\dot \CD$ that contains the edge $\{\u_1,\u_2\}$ while the second one
produces a diagram $\ddot \CD$ that does not contain this edge. 
We assume that $\dot \CD$ and $\ddot \CD$ have no vertices in common
and that the sub-walk $\dot\CW$ can get to the root vertex $\u_1$
along the edge $\{\u_1,\u_2\}$ only. 
Thus  by the recurrence the  diagrams $\dot \CD$ and $\ddot \CD$ are such that their
skeletons $\hat G(\dot \CD)$ and $\hat G(\ddot\CD)$ are   trees. We consider all possible 
sub-walks $\dot\CW$ and $\ddot \CW$ of this kind and therefore the set of walks of $k$ steps
$\bW_k$ that we construct by this recurrence procedure contains all possible 
tree-type walks. This procedure also allows one
to compute the total weight of walks from $\bW_k$.  
Now let us give rigorous definitions to the notions used above.

We denote by $\bW_k^{(r)}$ the family of such walks from $ \bW_k$
that have $r\ge 0$ steps either of the form $(\u_1,\u_i)$ or of the form
$[\u_1\u_i\u_1]$, in other words $r$ steps out of $\u_1$ and denote by 
$\Th(k,r)= \Th^{(v,\phi_1)}(k,r)$ the sum of the weights of all walks 
from $  \bW_k^{(r)}$,
$$
\Th(k,r)= \sum_{\CW\in \bW_k^{(r)}} \Pi(\CW).
$$
If $r=0$, then $\bW_k^{(0)}$ is non-empty if and only if $k=0$ and in this 
case $\bW_0^{(0)}$ contains one trivial walk of zero steps $\CW_0=(\u_1)$. Therefore we accept that 
 $\Th(k,0) = \delta_{k,0}$ for all $k\ge 0$.
 Let us also  introduce
 the set of walks $\dot  \bW_j^{(g)}$ of $j$ steps 
such that their trees  have only one edge $\{\u_1, \u_2\}$ 
attached to the root $\u_1$ and in $ \dot \CW_{j}^{(g)}$
 there are $g$ steps either of the form $(\u_1,\u_2)$
 or of the form $[\u_1\u_2\u_1]$. 
Now we can formulate the following statement that we refer to as the first decomposition lemma \cite{KSV,KV}.

\vskip 0.3cm
{\bf Lemma 2.2.} {\it For any $k\ge r\ge 1$
$$
\Th(k,r) = \sum_{g=1}^r \ \sum_{s=r-g}^{k-g} \ 
{r-1\choose g-1} \ \U(k-s,g) \, \Th(s,r-g), \quad k\ge  r\ge 1,
\eqno (2.33)
$$
where 
$$
\U(k-s,g) = \sum_{\CW\in \dot \CW_{k-s}^{(g)}} \Pi(\CW).
$$ 
 }

{\it Proof. } Given $g\le r$, 
let us consider a walk  $ \CW $ that performs $r$ steps out of the root $\u_1$
and $g$ steps of the form either $(\u_1, \u_2)$ or $[\u_1\u_2\u_1]$. 
If $g=r$, then by equality $\Th(s,0)= \d_{s,0}$ we have $s=0$ and 
(2.33) takes the form
$
\Th(k,r)=\U(k,r)$ that is true by the definition of $\U(k,r)$.

If $r-g>0$, then there exists at least one step of $\CW$ 
of the form either $(\u_1, \u_i)$ or $[\u_1\u_i\u_1]$ such that $\u_i\neq\u_2$. Denoting by $\dot  \t_1$
the instant of time of the first step of this kind,
we can consider the sub-walk $\dot \CW^{(1)}_{\ddot \t_1-1}=(\CW(1),\dots, \CW(\dot\t_1-1))$ 
of $\ddot \t_1-1$
steps. We say that $\dot \CW^{(1)}_{\ddot \t_1-1}$ is 
the partial sub-walk of the A-type. 
Regarding the run of the walk after the instant $\ddot\t_1-1$, we find 
the first instant of time $\dot \t_2>\ddot \t_1$  such that the step 
$(\CW(\dot\t_2-1),
\CW(\dot \t_2))$ is equal to either $(\u_1,\u_2)$ or $[\u_1\u_2\u_1]$.
We say that the sub-walk of $\CW$ given by the sequence
$(\CW(\ddot\t_1), \dots, \CW(\dot\t_2-1))=
\ddot 
\CW^{(1)}_{\dot {\t_2}-\ddot \t_2}$ is the partial sub-walk of the B-type.
Then we proceed by construction of the partial sub-walks $\dot \CW^{(l)}, l\ge 2$ and $
\ddot \CW^{(j)}, j\ge 2$, if they exist. 
Denoting by $s$ the total number of steps in the partial B-type sub-walks 
we can consider the following two new sub-walks of $\CW$
$$
\dot \CW_{k-s} = (\CW(1), \dots, \CW(\ddot\t_1-1), \CW(\dot \t_2),
\dots, \CW(\ddot \t_2-1), \dots )= \uplus_{l\ge 1} \dot \CW^{(l)} 
$$
that we refer to A-type sub-walk and B-type sub-walk of $\CW$, respectively.
and 
$$
\ddot \CW_s = \uplus_{j\ge 1} \ddot \CW^{(j)},
$$
where we do not indicate the number of steps in the A-type and B-type 
partial sub-walks in order not to complicate the denotations.
By the construction, $\dot \CW_{k-s}\in \check \bW_{k-s}^{(g)}$
while $\ddot \CW_s\in \bW_{s}^{(r-g)}$.

\vskip 0.3cm
Regarding the  walk $\CW_{12}$ (2.31), we see that the B-type sub-walk $\ddot \CW_3$
is given by one part of  three  steps,  $\ddot \CW_3=(\u_1, \u_4, [\u_5], \u_1)$. Here $\ddot \t^{(1)}= 4$ and $\dot \t^{(2)} = 7$.
The A-type sub-walk  of $\CW_{12}$ is given by the composition of two partial sub-walks;
the first one consists of three steps, $\dot \CW^{(1)}=(\u_1, \u_2, [\u_3],\u_1)$
and the second one has six steps, $ \dot \CW^{(2)}= (\u_1,\u_2, 
[\u_6],[\u_1], \u_3,\u_2, \u_1)$, so that 
$$
\dot \CW_9 = (\u_1, \u_2, [\u_3],\u_1,\u_2, 
[\u_6],[\u_1], \u_3,\u_2, \u_1).
$$
In this example, $r=3$, $g=2$ and $s=3$.

\vskip 0.3cm

Let us denote by  $\bW_k^{(r,g)}(s)$ the family of  walks 
$ \CW_k$ that perform $r$
steps out from the root $\u_1$ with $g$
steps of the form either $(\u_1,\u_2)$ or $[\u_1\u_2\u_1]$, $1\le g\le r$ and such that its B-type sub-walk $\ddot \CW_s$ consists of $s$ steps. 
By construction, $\CW_k $ uniquely determines its A-type sub-walk $\dot \CW_{k-s}$ and B-type sub-walk $\ddot \CW_s$. 
However, the pair $(\dot \CW_{k-s}, \ddot \CW_s)$ does not 
determines uniquely $\CW_k$ because when creating $\dot \CW_{k-s}$
we have lost the information about  the repartition of $\dot \CW_{k-s}$
into the partial sub-walks $\dot \CW^{(l)}$. The same is true for the
B-type sub-walk $\ddot \CW_s$. 
This means that to determine uniquely $\CW_k$
by its A-type sub-walk and B-type sub-walk, we have to 
know an additional information about the next in turn step
out from $\u_1$ - does it belong to the A-type partial sub-walk or to the 
B-type partial sub-walk.

The maximally possible number of partial 
sub-walks
of $\dot \CW_{k-s}$ is given by the number of the steps out from $\u_1$ and this is $g$. 
The maximally possible
number of partial sub-walks of $\ddot \CW_s$ is also equal to the number of the steps out from $\u_1$ and this is $r-g$. 
The first step out from $\u_1$
always belongs to the A-type partial sub-walk. 
Then we observe that there are ${r-1\choose g-1}$ possibilities to attribute 
$g-1$ labels "A" to $r-1$ steps out form $\u_1$, the remaining 
get the labels "B". Since the first and the second components of $(\dot \CW_{k-s}, \ddot \CW_s)$
can be independently chosen from 
$\dot \bW_{k-s}^{(g)}$ and $\bW_s^{(r-g)}$, respectively, we can write that 
$$
\#\bW_k^{(r,g)}(s) = {{r-1}\choose{g-1}}\times 
\# \dot \bW^{(g)}_{k-s} \times 
\# \bW^{(r-g)}_s.
$$ 
Taking into account that by definition $\CD(\dot \CW_{k-s})$ and 
$\CD(\ddot \CW_s)$ have no edges in common, we can write that 
$\Pi(\CW_k)= \Pi(\dot \CW_{k-s}) \Pi(\ddot \CW_s)$,
and therefore
$$
\sum_{\CW \in \bW_k^{(r,g)}(s)} \Pi(\CW)= 
{r-1\choose g-1} \times 
\sum_{\dot \CW\in \dot  \bW^{(g)}_{k-s}} 
\Pi(\dot \CW) \times \sum_{\ddot \CW \in  \in \bW^{(r-g)}_s)} 
\Pi{(\ddot \CW)}.
\eqno (2.34)
$$
Summing both parts of (2.34) over $r$, $g$ and $s$, we get equality (2.33).
 Lemma 2.2 is proved. $\Box$

\vskip 0.3cm
The second decomposition lemma concerns   \mbox{$\U(k-s,g)$} 
given by the total weight of walks
from the set $\dot \bW_{k-s}^{(g)}$. In what follows, $g$  represents   the sum
of $w\ge 0$ blue steps of the form $(\u_1,\u_2)$ 
and $f\ge 0$ red steps of the form $[\u_1\u_2\u_1]$
 performed
by a walk from $\dot \bW_{k-s}^{(g)}$, $g= w+f$. 

\vskip 0.3cm
{\bf Lemma 2.3.} {\it The following relation holds,
$$
\U(k-s,g) = \sum_{w=0}^g \  \sum_{h= 0}^{k-s-g}\ \ \sum_{t=0}^{k-s-g-w-h} \ 
{v^{2(g+h)}\over \phi_1^{g+h-1}}\,  
$$
$$
\times \ {g\choose w} \ {w+h-1\choose  h}^* \, \,   { w+h+t-1\choose t}^*\ \,  \Th(k-s-g-w-h, t),
\eqno (2.35)
$$
for any $g\ge 1$, $k-s\ge  g$. }

\vskip 0.3cm

{\it Proof.} Among  $g$ steps from  $\u_1$ to $\u_2$, 
we have to point out $w$ steps 
that will be the blue ones. This gives the first binomial coefficient of (2.35). 
Since each blue step 
is coupled with  the inverse blue step,
the walk has to perform $w$ blue steps of the form $(\u_2,\u_1)$. 
Assuming that the walk $\dot \CW_{k-s}^{(g)}$ makes $h$ red steps of the form $[\u_2\u_1\u_2]$ and taking
into account the fact that the last step from $\u_2$ to $\u_1$  is  the blue one,
we have to point out  among $w+h-1$ steps from $\u_2$ to $\u_1$  $w-1$ blue steps. 
This gives the second binomial coefficient from the right-hand side of (2.35).
Finally, assuming that the walk
$\dot  \CW_{k-s}^{(g)}$ performs $t$ steps of the form 
either $(\u_2, \u')$ or $[\u_2\u'\u_2]$ such that  $ \u'\neq \u_1$, 
we have to point out these $t$ steps among $w+h+t-1$ steps out from $\u_2$
performed by the walk. This gives the last binomial coefficient from the right-hand side of (2.35).

We see that in $\dot \CW_{k-s}^{(g)}$
the exists a number of partial sub-walks each of them starting and ending 
at $\u_2$ that do not contain the letter $\u_1$, either ordinary or generalized. Gathering these partial sub-walks into the one sub-walk
we get an element of $\bW_{k-s-g-w-h}^{(t)}$. We refer to this sub-walk as the
upper sub-walk of $\dot \CW_{k-s}^{(r)}$.
The total weight of the elements  of the set  $\bW_{k-s-g-w-h}^{(t)}$
is given by the factor
$\Th(k-s-g-w-h,t)$ standing in the right-hand side of (2.35).

Given $w$ and $h$, we see that the multi-edge ${\u_1,\u_2}$ of $\CD(\CW_k)$,
$\CW_K \in \dot \bW_{k-s}^{(g)}$ contains $2w$ blue edges and 
$g-w +h$ red edges. According to the definition of $\Pi(\CW)$, this multi-edge produces the factor 
$v^{2(g+h)}/\phi_1^{g+h-1}$ in (2.35).
\vskip 0.3cm 

To show that the lower limits of summation of (2.35) are correctly indicated, 
let  us  consider the contribution of the walks with $w=0$. 
This means
that the first edge from $\u_1$ to $\u_2$ of the corresponding diagram $\CD$ does not contain any blue edge 
and therefore there are no blue edges of the form $(\u_2,\u_1)$. 
The walk passes the edge $(\u_1,\u_2)$ $g$ times by the red edges and  the upper sub-walk  is empty. 
Regarding the right-hand side of (2.35) and remembering  property  (2.5),
we see that if $w=0$, then $h=0$ and $t=0$. 
Then equality   $\Th(k-s-g-w-h,0) = \d_{k-s-g, 0}$ 
implies  that $k-s-g=0$ and that $\U(g,g) = v^{2g}/\phi_1^{g-1}$. 
Lemma 2.3  is proved. $\Box$

\vskip 0.2cm
 
 Combining (2.33) and (2.35), we get relation (2.4). 
 It follows from the definition of $\Th(k,r)$  and (2.3) that
 $$
 \Pi(\hat \bW_k ) = \sum_{r=1}^k \Th(k,r) = m_k^{(v,\phi_1)}
 $$
 and then (2.32) implies (2.2).
This completes the proof of 
\mbox{Theorem 2.1.}

 \subsection{Limiting transition of infinite $\phi_1$}
 
  Let us  study $m_k^{(v,\phi_1)} $ (2.3) in 
   the limiting transition $\phi_1\to \infty$. 
Regarding  (2.4), we  observe  that the only term that  gives a non-vanishing contribution to the 
  the right-hand side of (2.4) corresponds to the case of $g+h=1$. 
  Since $g\ge 1$, then $h=0$ and $w$ can take only two values, zero and one. 
  Taking into account this observation and 
  denoting 
   $$
\th (k,r) = \lim_{\phi_1\to \infty} \Th^{(\phi_1)}(k,r),
 $$ 
 we deduce  from (2.35) relation
  $$
  \U(k-s,1) = v^2 \sum_{w=0,1}
\sum_{t=0}^{k-s-1-w} {1\choose w}\    
{w-1\choose 0}^*\  {w+t-1\choose t}^* \  \theta (k-s-1-w,t)
$$ 
  $$
  = v^2 \theta(k-s-1,0) +  v^2 \sum_{t=0}^{k-s-2} 
  \theta(k-s-2,t)
    = v^2 \delta_{k-s-1,0} +  v^2 \sum_{t=0}^{k-s-2} 
  \theta(k-s-2,t).
  $$
Substituting this expression  into (2.33),  we conclude that 
  $$
   \th (k,r) = v^2 \th(k-1,r-1) + v^2 \, \sum_{s=r-1}^{k-1} \ \th (s,r-1) 
  \sum_{t=0}^{k-s-2} \th (k-s-2, t) .
  \eqno (2.36)
  $$
  Denoting 
  $$
  \mu_k= \sum_{r=1}^k \th(k,r) 
  $$
  and changing the order of summation in the second term of (2.36),
 we get the following recurrence that determines $\{\mu_k\}_{k\ge 0}$,
 $$
 \mu_k = v^2 \mu_{k-1} + v^2 \sum_{j=0}^{k-2} \mu_j \, \mu_{k-j-2},
 \quad k\ge 2
 \eqno (2.37)
 $$
  with  obvious initial values $\mu_0 =1$ and $\mu_1 = v^2$. 
        Regarding the generating function 
    $$
    g(z) = -  \sum_{k=0}^\infty  \mu_k  {1\over z^{k+1}},
    $$
    one can  
  easily deduce from (2.37) relation
  $$
  g(z) = {1\over v^2 -z-v^2 g(z)}.
  \eqno (2.38)
  $$  
Equation (2.38) is uniquely solvable in the class of  functions $ g(z)$ such that 
$  \hbox{Im\,} z \cdot \hbox{Im\,} g(z) \ge 0 $      and therefore $g(z)$ represents the Stieltjes transform of the limiting measure
        $\tilde \s(\l)$ that is   the widely known Wigner semi-circle distribution \cite{W} shifted by $v^2$; 
$$
{d\tilde \s_v \over d\l}=  {1\over 2\pi v^2}
\begin{cases}
 \sqrt{4v^2 - (\l-v^2)^2}
  , & \text{if  $\vert  \l - v^2\vert \le 2\vert v\vert$}, \\
0, & \text {otherwise.}
\end{cases} 
  \eqno (2.39)
$$
This eigenvalue distribution has been already obtained as the eigenvalue  distribution of the
random matrices of the form (1.14) determined  for the Erd\H os-R\'enyi 
random graphs in the limit of the infinite average vertex degree, $\r\to\infty$ \cite{K-15}. 

\vskip 0.5cm
To complete this section, let us note that by repeating the proof of Theorem 2.1, it is not hard to  show that 
the moments of the normalized adjacency matrix $\tilde A_N^{(R)}$ (1.14) converge in the limit $(N,R)^\circ\to\infty$,
$$
\lim_{(N,R)^\circ\to\infty} \E \left( {1\over N} \Tr \left(\tilde A_N^{(R)}\right)^k\right) = \ell_k^{(v,\phi_1)}=
\begin{cases}
 L_p^{(v,\phi_1)}
  , &\text{if  $k = 2p$}, \\
0, &\text {if $k = 2p+1$}
\end{cases} 
\eqno (2.40)
 $$     
 for given $k=0,1,2,\dots$,  where the moments 
 $ L_p^{(v,\phi_1)}= \sum_{r=1}^p \L(p,r)$, $p\ge 0$ can be found from the following recurrent relations 
 \cite{KSV,KV},
 $$
 \L(p,r) = \sum_{g=1}^r \ {v^{2g}\over \phi_1^{g-1}}
 {r-1\choose g-1}
  \sum_{s=r-g}^{p-g}  \L(s,r-g)  \sum_{t=0}^{p-s-g} 
  {g+t-1 \choose t} \L(p-s-g,t),
  \eqno (2.41)
 $$
with the initial conditions $\L(p,0) = \d_{p,0}$, $p= 0,1,2,\dots$

Introducing the auxiliary numbers
$$
\fL^{(i)}_p = \sum_{r=0}^p \ {(r+1)(r+2)\cdots (r+i-1)\over 1\cdot 2\cdots (i-1)}
\L(p,r),
$$
where $ i= 1, \dots, p,
$
it is easy to deduce from  (2.41)  relation
$$
\fL^{(1)}_p =  v^2 \sum_{j=0}^{p-1} \ \fL^{(1)}_{p-1-j} \, \fL^{(1)}_{j} + 
{v^2\over \phi_1 } \sum_{j=0}^{p-2} \ \fL^{(2)}_{p-2-j} \, \fL^{(2)}_{j}
+\dots + {v^{2p}\over \phi_1^{p-1 }} \fL^{(p)}_0 \fL^{(p)}_0 . 
\eqno (2.42)
$$
It follows from (2.42) that 
$$
 \lim_{\phi_1\to\infty} \fL^{(1)}_p\to \CL_p(v),
 \eqno (2.43)
 $$ 
 where $\CL_p(v)$ verifies the famous
 recurrence
$$
\CL_p(v) = v^2\ \sum_{j=0}^{p-1} \ \CL_{p-1-j}(v)\, \CL_{j}(v),\quad \CL_0(v) = 1
\eqno (2.44)
$$ 
that determines the moments of the Wigner semi-circle distribution \cite{W},
$$
 \CL_p(v) =v^{2p} {(2p)!\over p!\, (p+1)!}, \quad p=0,1,2,\dots
 $$
Then (2.42) can be regarded as a generalization of (2.44) to the case
of the random graphs with the average vertex degree $\phi_1$ finite.

Taking into account  (2.42), we observe that the moments of $\tilde A_N^{(R)}$ (2.40) 
converge in the additional limiting transition  \mbox{$\phi_1\to\infty$}
$$
\lim_{\phi_1\to\infty} L^{(v,\phi_1)}_{2p}  = \CL_p(v)
\eqno (2.44)
$$
and therefore the limiting eigenvalue distribution
of $\tilde A_N^{(R)}$ weakly converges in average  
in the limit $(N,R)^\circ\to\infty $ (1.10) and $\phi_1\to\infty$ 
to the semi-circle distribution. 


\section{ Ihara zeta function of large random graphs}

Let us  discuss   our results in relations with 
the properties of Ihara zeta function and the graph theory Riemann Hypothesis.
In contrast with the previous sections, 
some statements are non-rigorous here in the sense that they 
are based on relations that are not proved yet. 
However, the  conjectures we formulate 
could be regarded as a source of new interesting questions of 
spectral theory of large random matrices.

\subsection{Weak convergence of eigenvalue distribution}

First of all, let us say 
that the limiting moments $m_k^{(v,\phi_1)}$ (2.3) verify the upper bound
$$
m_k^{(v,\phi_1)}\le (C_1v^2)^k k^{k+1}
\eqno (3.1)
$$
with a constant $C_1\ge C= C(v,\phi_1)$ (see Section 4 for the proof).  
The Carleman's condition (see e.g. monograph \cite{Akh}) is obviously satisfied,
$$
\sum_{l\ge  0} {1\over \left( m_{2l}^{(v,\phi_1)}\right)^{1/(2l)}} =+\infty
$$
 and  therefore the limiting measure $\sigma^{(v,\phi_1)}$ is uniquely determined
by the family $\{ m_k^{(v,\phi_1)}\}_{k\ge 0}$. 
Then we can say that  Theorem 2.1 implies the weak convergence 
in average of measures $\s_{N,R}$ to  $\sigma^{(v,\phi_1)}$. This means that for any continuous bounded function 
$f$: $\bR\mapsto \bR$ the following is true,
$$
\lim_{(N,R)^\circ \to\infty}\  \int_{\bR} f(\l ) \, d\bar \s_{N,R}(\l) = 
\int_{\bR} f(\l ) \, d \s^{(v,\phi_1)}(\l).
\eqno (3.2)
$$
Relation (1.13) can be rewritten as 
$$
\Psi_N^{(R)}(v) = {1\over N}
\log \det \left(\left( 1-{v^2\over \phi_1 } \right)I + H_N^{(R)}(v)\right)
$$
$$
= \int_{\bR} \log (1_{\phi_1} + \lambda) \, d \s_{N,R}(\l),
\eqno (3.3)
$$
where we denoted $1_{\phi_1} = 1 - v^2/\phi_1$.  
By this relation, the problem of convergence of IZF for a sequence 
of infinitely increasing graphs can be reduced to the analysis
of convergence of the spectral measures of operators $H^{(R)}_N$. 
This approach to the studies of IZF has been proposed for the first time 
in paper \cite{GZ}.

Taking into account the weak convergence of measures  (3.2), 
one could expect that the mathematical expectation 
of (3.3) $\E \Psi_N^{(R)}(v/\sqrt{ \phi_1})$, $v\in \fI$ exists for 
 some non-zero interval 
 $\fI\subset \mathbb R$ and  converges in average to the integral
over $d \s^{(v,\phi_1)}$ (3.2),
$$
 \E \Psi_N^{(R)}(v/\sqrt{\phi_1}) \to   \int_{\bR} \log (1_{\phi_1} + \lambda) \, d\s^{(v,\phi_1)}(\l),
\quad {\hbox{as}} \quad (N,R)^\circ\to\infty
\eqno (3.4)
$$
 and then
 $$
 \lim_{(N,R)^\circ\to\infty} {1\over N} \E \log Z_{\G_N^{(R)}}(v/\sqrt \phi_1) 
 = 
 - \left( {\phi_1\over 2}-1\right) \log \left(1- {v^2\over \phi_1} \right) -
 \Psi( v/\sqrt {\phi_1}),
\eqno(3.5) 
 $$
 where we denoted 
  $$
 \Psi( v/\sqrt {\phi_1})
=    \int_{\bR} \log (1+\l) \, d \s^{(v,\phi_1)}(\l).
$$
\vskip 0.2cm
Although convergence (3.4) is not proved, let us note that  accepting that the limiting transition $\phi_1\to\infty$ can be performed in (3.5)
and assuming that the limiting equality holds, 
one could expect that
$$
\lim_{\phi_1\to\infty} \lim_{(N,R)^\circ\to\infty} {1\over N} \E \log Z_{\G_N^{(R)}}(v/\sqrt \phi_1) 
= 
 {v^2\over 2} - \Upsilon(v),
 \eqno (3.6)
 $$
where 
 $$
\Upsilon(v) =  {1\over 2\pi v^2} \int_{-2\vert v\vert+v^2 }^{2\vert v\vert +v^2}  \log (1+\l) \, \sqrt{4v^2 - (\l-v^2)^2} d \l
$$
and therefore one could conclude that 
$$
\lim_{\phi_1\to\infty} \lim_{(N,R)^\circ\to\infty} {1\over N} \E \log Z_{\G_N^{(R)}}({v\over \sqrt \phi_1}) 
= \CF(v),
\eqno (3.7)
$$
where 
$$
\CF(v)=  {v^2\over 2}- {2\over \pi} \int_{-1}^1 \log (1+v^2 +2v \tau) \sqrt{1-\tau^2} \, d\tau.
\eqno (3.8)
$$
It is easy to see that the last integral exists for any real $v$ and that the limiting function is continuous for all $v\in \bR$.

\subsection{ Graph theory Riemann Hypothesis}

\noindent Regarding the family of finite regular graphs
with the vertex degree $q+1$ (in other terms, $(q+1)$-regular graphs), the graph theory analog of the Riemann Hypothesis can be formulated as follows \cite{HST,Terras-10}:
$$
Z_\G(q^{-s}) \ \ {\hbox {has no poles with $0<$ Re\,$s<1$ unless 
Re\,$s=1/2$.}}
\eqno (3.8)
$$

It can be shown that (3.8) is equivalent to the condition for the graph $\G$ to be the Ramanujan one \cite{LPS}; more precisely, if a connected regular graph $\G$ verifies (3.8), then \cite{HST}
$$
\rho_\G'\le 2 \sqrt{\rho_\G-1},
\eqno (3.9)
$$
where 
$$
\rho_\G = \max \{ \vert \l\vert: \ \l\in {\hbox {Sp}}(A_\G)\}
\ \  {\hbox{and }} \ \ 
\rho'_\G = \max \{ \vert \l\vert: \ \l\in {\hbox {Sp}}(A_\G), \vert \l\vert\neq \rho_\G\}
$$ 
and where we denoted by $\hbox{Sp}(A_\G)$ the ensemble of 
eigenvalues  of the 
adjacency matrix $A_\G$ of the graph $\G$. For any $(q+1)$-regular graph $\rho_\G= q+1$, so condition (3.9) reads as $\rho'_\G \le 2\sqrt{q}$. 

Numerical simulations of  \cite{MN} lead to the conjecture that the proportion 
 of regular graphs exactly satisfying the graph theory Riemann Hypothesis approaches $27\%$ as 
the number of vertices $N=\#V(\G)$ infinitely increases. 
From another hand, the Alon conjecture for regular graphs
proved by \mbox{J. Fridman} (see \cite{F-08} and also \cite{Bor}) says that
for any positive $\vep>0$ 
$$
{1\over \#\CG_N(q+1)}
\#\left\{\G:\  \G\in \CG_N(q+1),  \
\rho'_\G\ge 2\sqrt{q} +\vep
\right\}\to 0, \quad N\to\infty,
\eqno (3.10)
$$
where $\CG_N(q+1)$ is the family of all $(q+1)$-regular graphs
with the set of vertices $\{1,2, \dots, N\}$. This result means that 
 the Riemann Hypothesis (3.8) is ``approximately true'' for 
most of regular graphs of high dimension \cite{HST}.

 \vskip 0.2cm 
It can be shown that 
 (3.8)  is equivalent
to the statement that for finite connected $(q+1)$-regular graph $\G$
with $q\ge 2$
$$
Z_\G(u) \ {\hbox{is pole free for}} \  {1\over q} < \vert u\vert <{1\over \sqrt q}.
\eqno (3.11)
$$
For the family of irregular graphs, one can formulate the weak graph theory Riemann Hypothesis saying that \cite{HST,ST-III}
$$
Z_\G(u) \ {\hbox{is pole free for}} \  R_\G < \vert u\vert <{1\over \sqrt{d_{\max}-1}},
\eqno (3.12)
$$
where $R_\G$ is the  radius of the largest circle of convergence of the Ihara zeta function $Z_\G(u)$ and $d_{\max}= d_{\max}(\G)$ is the maximum degree of $\G$.

\vskip 0.2cm

Let us return to the Ihara zeta function 
$Z_{\Gamma_N^{(R)}}$ of random graphs $\Gamma^{(R)}_N$ (1.8)
and consider the conjectured limiting expression for its normalized 
logarithm (3.7). It follows from the classical theorems that the limiting function 
$\CF(v)$ (3.8)  
 is continuous for all $v\in {\mathbb R}$ and therefore has no real poles. Moreover, the integral expression of the right-hand side of (3.8)
 can be continued to a function holomorphic in any domain
 $$
 C_{\epsilon} = \{v:\, v\in {\mathbb C}, \vert v\vert <1-\epsilon\},
 \ 0< \epsilon < 1.
 \eqno (3.14)
 $$
 From another hand, for any $v'\in {\mathbb C}$ such that 
 $\vert v'\vert=1$ and $v'\neq \pm 1$, there exists 
$\tau'$ such that $\log (1+v^2 +2\tau v)\sqrt{1-\tau^2}$ has a non-integrable singularity at this point. 

Remembering pre-limiting relation $v= u \sqrt {\phi_1}$ (1.12), we could 
deduce from (3.14)  that the function $\Psi(u) = \Psi(v/\sqrt{\phi_1})$  (3.5) has 
no real poles and no complex poles $u'$ such that 
$$
\vert u'\vert <{1\over \sqrt{\phi_1}},
\eqno (3.15)
$$ 
but has the  poles lying everywhere on the circle $\vert u\vert =1/\sqrt{\phi_1}$, $u\in \mathbb C$ excepting the regular points 
$-1/\sqrt{\phi_1}$ and $1/\sqrt{\phi_1}$.

Now we can formulate a conjecture
that the normalized logarithm of the Ihara zeta function of random graphs 
$Z_{\Gamma_N^{(R)}}(u)$ determined by (1.4) with 
the spectral parameter renormalized by the average vertex degree, 
$u = v/\sqrt {\phi_1}$,
converges in average  in the limit  $N\to \infty$, $R\to\infty$ and 
\mbox{$\phi_1\to\infty$} (1.6)
to a function  $\CF(v)$ (3.8),
$$
-{1\over N} \E \log Z_\G \left( {v\over 
\sqrt{
\phi_1}}\right) \to \CF(v);
\eqno (3.16)
$$
 the limiting function $\CF(v)$   has no poles inside the unit circle \mbox{$\{ v\in \mathbb C: \vert v\vert <1\}$.}
 This proposition is in agreement with (3.12) with the  difference that 
 the value $d_{\max}-1$ for non-random graphs is replaced by the averaged
 vertex degree
$ d_{\hbox{\begin{footnotesize}{av}\end{footnotesize}}}=\phi_1$
of random graphs. 
    Therefore the  statement (3.16) can be viewed as one more 
relaxed version 
 of the graph theory Riemann Hypothesis  for the ensemble of infinite 
 random graphs $\{\G_N^{(R)}\}$.
 
A conjecture similar to (3.16) can be put forward for the ensemble
 of  Erd\H os-R\'enyi $N$-dimensional graphs 
with the edge probability $\rho/N$. In this case the averaged vertex degree 
$d_{\hbox{\begin{footnotesize}{av}\end{footnotesize}}}=\phi_1$ in (1.14) and (3.16)
is replaced by  $\rho$ and the limiting transition considered 
is given by $N,\rho\to\infty$, \mbox{$\rho= o(N)$ \cite{K-15}.}

 \vskip 0.2cm

 Finally, let us stress that
 convergence (3.4) and the limiting transition (3.6)  for real
 and for complex $v$
 are assumed but not proved rigorously neither in the present  paper  nor in \cite{K-15}.
These interesting  questions remain open ones that would require deep studies of
fine spectral properties of the random matrix ensembles, such as the asymptotic behavior 
of the maximal eigenvalue of large random matrices $A_N^{(R,\phi)}$ (1.4).  
In these studies,  additional restrictions could be imposed; these restrictions
could be related  with the 
ratio between $N,R$ and $\phi_1$  for the ensemble
$H_N^{(R)}(v)$ and between $\rho$ and $N$ for the random matrix 
of \cite{K-15}, where the value $\rho= \log N$ is shown to be 
critical for the asymptotic behavior of the spectral norm (see 
 \cite{K-16} for more details). 

\section{Appendix: Proof of the upper bound (3.1)}

In this section we will show  that the moments $\{m_k^{(v,\phi_1)}\}$ (2.3) 
and $\{L_k^{(v,\phi_1)}\}$ (2.40)
verify the Carleman's condition.  
 \vskip 0.2cm 
 {\bf Lemma 4.1} {\it Let  $C$ be a  constant such that 
 $$
 C\ge 1 \quad {\hbox{and}} \quad C{\phi_1} \ge 1.
 \eqno (4.1)  
$$
Then for any $p\ge 1$, the following upper bound is true,
$$
\max_{1\le r\le p}  \L(p ,r) \le \left(Cv^2\right)^p \,p^{2p}.
\eqno (4.2)
$$
}
\vskip 0.2cm
{\it Proof.} We prove (4.2) by recurrence. Assuming that all terms $ \L(i,j)$ of the right-hand side
of (2.41) verify (4.2), we deduce from (2.41) that
$$
 \L(p,r) \le {\left(Cv^2\right)^{p}\over C}
  \sum_{g=1}^r \ 
 {1\over (C\phi_1)^{g-1}} {r-1\choose g-1} (p-g)^{2(p-g)}
  \sum_{s=r-g}^{p-g}  \ 
  {p-s \choose g} ,
  \eqno(4.3)
  $$
where we have used an obvious inequality 
$s^{2s} (p-s-g)^{2(p-s-g)}\le  (p-g)^{2(p-g)}$ and the well-known identity
$$
\sum_{l=0}^j {l+i-1\choose l} = {i+j\choose i}, \quad i\ge 1.
\eqno(4.4)
$$
Changing variables in the last sum of (4.3), using a version of (4.4)
$$
\sum_{i=0}^{p-r}{g+i\choose i}= \sum_{i=0}^{p-r}{g+i\choose g} = {g+p-r+1\choose g+1},
$$
and taking into account  (4.1),
we can write that 
$$
 \L(p,r) \le 
\left(Cv^2\right)^p   \sum_{g=1}^r 
 {r-1\choose g-1} (p-g)^{2(p-g)}(p-r+1)^{g+1}
$$
$$
 \le  \left(Cv^2\right)^p p^{p+1} 
  \sum_{g=1}^r \ 
 {r-1\choose g-1} (p-g)^{p-g}
$$
$$
 \le 
 \left(Cv^2\right)^p p^{p+1}(p-1)^{p-r}
 \sum_{g'=0}^{r-1} {r-1\choose g'}(p-1-g')^{r-1-g'} 
 \le \left(Cv^2\right)^p p^{2p}.
 $$
 Lemma 4.1 is proved. $\Box$

It follows from (4.2) that 
$
L_p^{(v,\phi_1)} \le p^{2p+1}
$
and  the series $\sum_{p\ge 0} (L_p^{(v,\phi_1)})^{-1/(2p)}$ diverges. 
Thus the family of moments $\left\{ \ell_k^{(v,\phi_1)}\right\}_{k\ge 0}$
(2.40) verifies the Carleman's condition (see e.g. monograph \cite{Akh}).

 \vskip 0.3cm 
 {\bf Lemma 4.2}
 {\it The following upper bound  
 $$
 \max_{1\le r\le k} \tilde \Theta(k,r)\le (Cv^2 k)^k, \quad k\ge 1
 \eqno (4.5)
 $$
for any positive $C$
 such that }
$$
{1\over C}\left( 1+ {1\over Cv^2}\right)  \exp\left\{{1\over C\phi_1}\right\}\le 1 \quad 
{{and}} \quad {1\over C\phi_1} \left( 1+ {1\over Cv^2}\right) \le 1.
\eqno (4.6)
$$

\vskip 0.2cm
{\bf Proof.} Assuming that all terms of the right-hand side of (2.4) 
verify (4.6), we get inequality
$$
\Th(k,r) \le 
  \sum_{g=1}^r\ \sum_{s=r-g}^{k-g} 
  {r-1\choose g-1}\ 
 \sum_{w=0}^g\ {g\choose w}   \sum_{h=0}^{k-s-g-w}\ \ 
 {v^{2(g+h)}\over \phi_1^{g+h-1}}
$$
$$
\times   \ {w+h-1\choose  h}^* \  { k-s-g\choose w+h} \ 
  \left(Cv^2(k-g-w-h)\right)^{k-g-w-h },
  \eqno (4.7)
 $$
where we have used an obvious generalization of (4.4),
$$
\sum_{l=0}^j {l+i-1\choose l}^* = {i+j\choose i}, \quad i\ge 0.
$$
Taking into account that $(k-g-w-h)^{k-g-w-h}\le (k-g)^{k-g-w-h}$
and using elementary inequalities 
$$
{k-s-g\choose w+h}\le {(k-g)^{w+h}\over (w+h)!}, \quad w\ge 0,\, h\ge 0
$$
and 
$$
{w+h-1\choose h}^* {1\over (w+h)!} \le {1\over h!},\quad w\ge 0, 
\, h\ge 0,
$$
we can write  that 
$$
 \sum_{w=0}^g\ {g\choose w}  \sum_{h=0}^{k-s-g-w}\ \ 
  {v^{2(g+h)}\over \phi_1^{g+h-1}}
 {w+h-1\choose  h}^*   { k-s-g\choose w+h} \ 
  \left(Cv^2(k-g)\right)^{k-g-w-h } 
  $$
  $$
\le \left(Cv^2(k-g)\right)^{k-g} \ \sum_{w=0}^g {g\choose w} 
\sum_{h=0}^{k-s-g-w}{1\over h!}\cdot  {v^{2(g+h)}\over \phi_1^{g+h-1}}
\cdot
{1\over \left(Cv^2\right)^{w+h}}  
$$
$$
\le 
 {v^{2g}\over \phi_1^{g-1}}
 \left(Cv^2(k-g)\right)^{k-g\, } e^{1/(C\phi_1)} 
 \left(1+ {1\over Cv^2}\right)^g .
 \eqno (4.8)
   $$
Returning to (4.7) and replacing there the sum over $s$  by $k-g+1$, we  get with the help of (4.8) the following inequalities,
$$
\Th(k,r) \le 
e^{1/(C\phi_1)} {\left(Cv^2\right)^k \over C} 
$$
$$
\times 
\sum_{g=1}^r {r-1\choose g-1} {1\over \left( C\phi_1\right)^{g-1}}
 (k-g+1) (k-g)^{k-g} \left( 1+ {1\over Cv^2}\right)^g 
$$
$$
\le {\left(Cv^2\right)^k } e^{1/(C\phi_1)}  {1\over C}  
\left( 1+ {1\over Cv^2}\right) k
$$
$$
\times (k-1)^{k-r}
\sum_{g'=0}^{r-1} {r-1\choose g'} 
  (k-1)^{r-1-g'} \left( {1\over C\phi_1}\left(1+ {1\over Cv^2}\right)\right)^{g'} 
$$
$$
\le {\left(Cv^2\right)^k } e^{1/(C\phi_1)}  {1\over C}  
\left( 1+ {1\over Cv^2}\right) k^{k-r+1}
\left( k-1 + {1\over C\phi_1}\left(1+ {1\over Cv^2}\right)\right)^{r-1}.
$$
Taking into account (4.6), we get (4.5). Lemma 4.2 is proved. $\Box$

\vskip 0.2cm
It follows from (4.5) that the upper bound (3.1) is true 
and therefore the family of moments 
$\{ m_k^{(v,\phi_1)}\}_{k\ge 0}$ (2.2)  verifies the Carleman's condition
\cite{Akh}.

\end{document}